\documentclass[amsmath,amssymb,twocolumn,prd,preprintnumbers,showpacs]{revtex4-2}

\usepackage{hyperref}
\usepackage{slashed}
\usepackage{graphicx}
\usepackage{amsmath}
\usepackage{amssymb}
\usepackage{graphicx}
\usepackage{xcolor}
\usepackage[caption=false]{subfig}
\usepackage{soul}

\begin{document}

\title{Cosmology in the presence of diffeomorphism-violating, \\ nondynamical background fields}

\author{Carlos M. Reyes$^{1}$}
\email[Electronic mail:]{creyes@ubiobio.cl}
\author{Marco Schreck$^{2}$}
\email[Electronic mail:]{marco.schreck@ufma.br}
\author{Alex Soto$^{3}$}
\email[Electronic mail:]{alex.soto@ncl.ac.uk}
\affiliation{$^{1}$ Centro de Ciencias Exactas, Universidad del B\'{\i}o B\'{\i}o, Casilla 447, Chill\'{a}n, Chile}
\affiliation{$^{2}$ Departamento de F\'{\i}sica, Universidade Federal do Maranh\~{a}o, \\
Campus Universit\'{a}rio do Bacanga, S\~{a}o Lu\'{\i}s (MA), 65080-805, Brazil}
\affiliation{$^{3}$ School of Mathematics, Statistics and Physics, Newcastle University, Newcastle upon Tyne, NE1 7RU, UK}

\begin{abstract}

We consider diffeomorphism violation, which is parameterized by nondynamical background fields of the gravitational Standard-Model Extension (SME), and study its effects on the time evolution of the Universe. Our goal is to identify background field configurations that imply stages of accelerated expansion without exotic forms of matter and radiation present. Although our approach gives rise to a set of restrictive conditions, configurations are encountered that exhibit this property or show other interesting behaviors. The findings of our article, which is among the first to apply the SME to a cosmological setting, provide an initial understanding of how to technically incorporate background fields into the cosmological evolution equations and what their phenomenological impact may be.

\end{abstract}
\maketitle

\section{Introduction}

Luminosity distances of type Ia supernovae confirm that the current expansion of the Universe is accelerating~\cite{SupernovaSearchTeam:1998fmf,SupernovaCosmologyProject:1998vns,SupernovaCosmologyProject:2008ojh,Pan-STARRS1:2017jku}. Scans of temperature fluctuations in the cosmic microwave background recorded by WMAP~\cite{WMAP:2003elm,WMAP:2008lyn} and Planck~\cite{Planck:2018vyg}, the detection of baryon acoustic oscillations~\cite{SDSS:2005xqv,eBOSS:2020yzd} as well as the Dark Energy Survey~\cite{DES:2021wwk} further corroborate this finding. However, the nature of the fundamental physics responsible for this acceleration remains a complete mystery. Thus, Turner~\cite{Turner:1998mg} coined the term Dark Energy to refer to its exotic and completely unknown character. Studying the dynamics of mechanisms that imply a stage of accelerated expansion is a hot topic in cosmology.

The standard picture, called $\Lambda$CDM, describes Dark Energy via a cosmological constant $\Lambda$, which Einstein first of all introduced to keep the Universe static \cite{Einstein:1917}. Although cosmological data strongly disfavor a steady-state Universe, a nonzero cosmological constant has found a resurgence. After all, a positive value of $\Lambda$ is associated with a constant, homogeneous energy density permeating space, which could, indeed, drive a stage of accelerated expansion of the Universe. Assuming that the $\Lambda$CDM model is correct \cite{Lonappan:2017lzt}, Dark Energy would contribute around 68\% to the overall energy content of the Universe~\cite{Planck:2018vyg}. Although the $\Lambda$CDM model describes the current observations well, the smallness of the cosmological constant \cite{Carroll:2000fy} is still a conundrum \cite{Weinberg:1988cp}. On the one hand, alternatives have been proposed to give a satisfactory explanation of the nature of Dark Energy without introducing a cosmological constant such as the phantom model~\cite{Caldwell:1999ew,Caldwell:2003vq}, quintessence~\cite{Ratra:1987rm,Wetterich:1987fm,Wetterich:1994bg,Caldwell:1997ii,Hebecker:2000zb}, and $k$-essence~\cite{Armendariz-Picon:1999hyi,Armendariz-Picon:2000nqq,Armendariz-Picon:2000ulo,Mukherjee:2018}. Furthermore, there are approaches, e.g., q-theory \cite{Klinkhamer:2009nn,Klinkhamer:2009gm,Klinkhamer:2016iaw}, whose objective is to provide an explanation for the actual minuscule value of the cosmological constant.

On the other hand, an accelerated expansion of the Universe is not only a matter of the current epoch. An early expansion is believed to have occurred in a process called inflation, which solves issues of the standard Big Bang such as the horizon and flatness problems~\cite{Guth:1980zm,Starobinsky:1980te,Albrecht:1982wi,Linde:1983gd} (see, e.g., Refs.~\cite{deHaro:2022jrd,Cotsakis:2022xgp,Schmitz:2022hsz} for recent reviews on these particular issues, amongst other prominent topics in cosmology). As in the case of Dark Energy, there is no consensus on the exact model that explains this phenomenon, but they vary in the shape of the potential as well as the framework they are embedded in. However, the common feature is the presence of a scalar field within a slow-roll regime. The reader may wish to consult Ref.~\cite{Baumann:2009ds}, which provides an excellent review focusing on the basics of inflation.

Therefore, the exact root cause of stages where the expansion of the Universe is accelerated remains an open problem. Specific models are usually constructed by incorporating exotic sources of matter and radiation in addition to baryonic matter and electromagnetic radiation. Our interest in this paper is to generate an accelerated stage with only standard matter and radiation present. To do so, we modify General Relativity (GR) by introducing nondynamical background fields that violate diffeomorphism invariance.

Diffeomorphism invariance is the symmetry behind the dynamical spacetime structure incorporated in GR, which describes the gravitational laws of nature predominantly important at macroscopic length scales. Candidate fundamental theories such as string theory~\cite{Kostelecky:1988zi,Kostelecky:1989jp,Kostelecky:1989jw,Kostelecky:1991ak,Kostelecky:1994rn} and loop quantum gravity~\cite{Gambini:1998it,Bojowald:2004bb} as well as noncommutative spacetime geometry~\cite{AmelinoCamelia:1999pm,Carroll:2001ws,Bailey:2018ifc}, spacetime foam~\cite{Klinkhamer:2003ec,Bernadotte:2006ya,Hossenfelder:2014hha}, the implementation of nontrivial spacetime topologies~\cite{Klinkhamer:1998fa,Klinkhamer:1999zh,Klinkhamer:2002mj,Ghosh:2017iat}, and UV completions of GR, e.g., Ho\v{r}ava-Lifshitz gravity~\cite{Horava:2009uw,Bluhm:2019ato} have been demonstrated to imply a breakdown of Lorentz invariance. A fundamental energy scale is associated with Lorentz symmetry violation and the latter usually coincides with the Planck energy. Such effects would be strongly suppressed at much lower energies, but are still likely to leave fingerprints. In the presence of a gravitational field, the counterpart of (global) Lorentz violation is a breakdown of the fundamental symmetries of GR, i.e., local Lorentz symmetry on the one hand and diffeomorphism invariance on the other. Our focus is on the latter.

Any departure from diffeomorphism symmetry is expected to have strong consequences on the time evolution of the Universe. From a formal point of view, the Hamiltonian formulation \cite{Arnowitt:1962hi,Hanson:1976,Poisson:2002,Poisson:2004,Bertschinger:2005,Arnowitt:2008} reveals that diffeomorphism violation may alter the constraint structure and algebra of GR \cite{Gambini:1996}. As a part of searches for Planck scale effects in gravity the possibility of background fields breaking diffeomorphism symmetry has been considered within an effective field theory framework known as the gravitational Standard-Model Extension (SME)~\cite{Kostelecky:2003fs,Bailey:2006fd,Kostelecky:2017zob,Mewes:2019dhj,Kostelecky:2020hbb}. First bounds on diffeomorphism violation via modified, nondynamical spin-gravity couplings were determined in the recent papers \cite{Kostelecky:2021tdf,Ivanov:2021bvk}. The gravitational SME provides one powerful branch among the numerous possibilities of modifying GR that are on the market \cite{Will:2014kxa,Heisenberg:2018vsk,Tasson:2016xib,Petrov:2020,Shankaranarayanan:2022wbx,Mariz:2022oib}. It covers particular modified-gravity theories such as Brans-Dicke theory \cite{Brans:1961sx} and dRGT massive gravity \cite{deRham:2010kj} where the connection to the latter was established in Ref.~\cite{Bluhm:2019ato}.

The Arnowitt-Deser-Misner (ADM) decomposition of spacetime~\cite{Arnowitt:1962hi,Misner:1973,Arnowitt:2008} serves as the formal foundation for establishing the Hamiltonian formulation of a modified-gravity theory based on the gravitational SME~\cite{ONeal-Ault:2020ebv,Reyes:2021cpx,Reyes:2022mvm}. Our recent results presented in Ref.~\cite{Reyes:2022mvm} show that the Hamiltonian formulation of the gravitational SME, whose construction we focused on in Ref.~\cite{Reyes:2021cpx}, is consistent with the covariant approach developed earlier in Refs.~\cite{Kostelecky:2003fs,Bailey:2006fd}. This outcome provides a strong foundation that the current paper can rest on, since we will also make use of the ADM formalism.

The objective of this new analysis is to investigate the time evolution of the Universe in the presence of the nondynamical background fields considered in Refs.~\cite{Reyes:2021cpx,Reyes:2022mvm}. In particular, we are interested in constructing scenarios of accelerated expansion without the need of exotic matter or radiation.
A crucial feature of our approach is that it is perturbatively connected to the standard scenario, i.e., setting all background fields to zero reproduces standard cosmology. To the best of our knowledge, only a handful of papers have been written on the physics of the gravitational SME at cosmological scales, such as Refs.~\cite{Bonder:2017dpb,ONeal-Ault:2020ebv,Nilsson:2022mzq}. Thus, the purpose of the present work is to complement the sparse exploration of this interesting topic carried out until now.

Our paper is structured as follows. Section \ref{sec:sectionII} introduces the model, discusses its theoretical foundations, and sets the base for studying modified cosmologies in the remainder of the text. Section \ref{sec:cosmo} describes how to derive the first and second modified Friedmann equations from the ADM-decomposed action of the model as well as the modified Einstein equations. We are doing so for all background field configurations simultaneously. To simplify the first part of the analysis, the background fields are assumed to be static, i.e., unaffected by the cosmological expansion itself. In Sec.~\ref{sec:inflation} we search for specific background field configurations that lead to stages of accelerated expansion. Standard matter and radiation only are taken into consideration. Here, we analyze each type of background field separately. Section~\ref{sec:relaxu} is dedicated to the generic case of time-dependent background fields, which is an extension of the computations performed in Sec.~\ref{sec:inflation}. Finally, our findings are concluded on in Sec.~\ref{sec:con}. Appendix~\ref{sec:alternative-equations} is devoted to deriving an alternative version of the second modified Friedmann equation by different means. Although the latter is not made use of in the main body of the paper, it is presented and commented on for completeness. Natural units are employed with $\hbar=c=8\pi G_N=1$ unless otherwise stated. Furthermore, the metric signature is $(-,+,+,+)$. The \textit{Mathematica} packages \textit{xTensor}~\cite{xTensor:2020} and \textit{OGRe}~\cite{Shoshany2021_OGRe} turned out to be highly beneficial for symbolic and explicit computations involving tensors and covariant derivatives in curved spacetimes.

\section{Model and characteristics}\label{sec:sectionII}

We consider a modification of the Einstein-Hilbert action with a cosmological constant $\Lambda$ \cite{Kostelecky:2003fs,Kostelecky:2020hbb}:
\begin{subequations}
\label{eq:action-initial}
\begin{align}
S&=\int_{\mathcal{M}} \mathrm{d}^4x\,(\mathcal{L}^{(0)}+\mathcal{L}_{\mathrm{SME}})+S_m\,, \displaybreak[0]\\[2ex]
\mathcal{L}^{(0)}&=\frac{\sqrt{-g}}{2}({}^{(4)}R-2\Lambda)\,, \displaybreak[0]\\[2ex]
\mathcal{L}_{\mathrm{SME}}&=\frac{\sqrt{-g}}{2}(-u {}^{(4)}R + s^{\mu \nu} {}^{(4)}R^T_{\mu \nu}+t^{\mu \nu \rho \sigma} {}^{(4)}C_{\mu \nu \rho \sigma})\,,
\end{align}
\end{subequations}
with the four-dimensional spacetime metric $g_{\mu\nu}$ of the spacetime manifold $\mathcal{M}$ where $g:=\det(g_{\mu\nu})$. The traceless Ricci tensor is denoted as ${}^{(4)}R_{\mu\nu}^T$, the Ricci scalar is defined by ${}^{(4)}R:= {}^{(4)}R^{\mu}_{\phantom{\mu}\mu}$, as usual, and ${}^{(4)}C_{\mu\nu\rho\sigma}$ constitutes the Weyl tensor. Furthermore, $u$ is a scalar-valued and $s^{\mu\nu}$, $t^{\mu\nu\varrho\sigma}$ are tensor-valued nondynamical background fields depending on the spacetime coordinates. The part $S_m$ corresponds to the matter action, which remains unspecified at this point.

The property of the background fields $u$, $s^{\mu\nu}$, and $t^{\mu\nu\varrho\sigma}$ being both nondynamical and coordinate-dependent implies a breakdown of diffeomorphism invariance. Inferred background fields that carry local-coordinate indices can be defined by transforming $u$, $s^{\mu\nu}$, and $t^{\mu\nu\varrho\sigma}$ to a local inertial reference frame via a background vierbein~\cite{Kostelecky:2020hbb}. The latter give rise to preferred directions in freely falling inertial frames, which implies local Lorentz violation, too.

Several papers such as Refs.~\cite{Bonder:2015maa,Bonder:2020fpn} clarified that the physics related to the fourth-rank tensor background $t^{\mu\nu\varrho\sigma}$ is involved. Hence, we will be working in a setting where the latter is discarded, such that our final model is described by the action
\begin{equation}
\label{eq:actionus}
S = \int_{\mathcal{M}} \mathrm{d}^4x\,\frac{\sqrt{-g}}{2}\left[(1 - u) {}^{(4)}R + s^{\mu \nu} {}^{(4)}R_{\mu\nu}-2\Lambda\right] +S_m\,.
\end{equation}
In comparison to Eq.~\eqref{eq:action-initial}, the trace of the Ricci tensor is kept for simplicity. In the remainder of the paper, the consequences of diffeomorphism symmetry violation parameterized by $u$, $s^{\mu\nu}$ on cosmological evolution ought to be studied. Performing the ADM decomposition of the spacetime manifold $\mathcal{M}$ is valuable in this context. The infinitesimal path length interval squared is then cast into the form
\begin{equation}
\label{eq:admline}
\mathrm{d} s^2 = - (N^2 - N_i N^i) \mathrm{d} t^2 + 2 N_i \mathrm{d} x^i \mathrm{d} t + h_{ij} \mathrm{d} x^i \mathrm{d} x^j\,,
\end{equation}
where $N$ is the lapse function, $N^{a}$ denote the shift vector components, and $h_{a b}$ is the induced metric in a three-dimensional hypersurface $\Sigma_t$ of the foliation. In the following, we will also employ the projector $h^{\mu}_{\phantom{\mu}\nu}$ into $\Sigma_t$ as well as the vector $n^{\mu}=(1/N,-N^i/N)$
orthogonal to $\Sigma_t$ and the associated covector $n_{\mu}=(-N,0,0,0)$, which are frequently in use within the ADM formalism. We can then decompose the action of Eq.~\eqref{eq:actionus} as
\begin{subequations}
\label{eq:ADM-action}
\begin{align}
S&=\int_{\mathcal M}\mathrm{d}t\mathrm{d}^3x\,\mathcal{L}_{\mathrm{ADM}}+S_m\,, \displaybreak[0]\\[2ex]
\label{eq:ADM-lagrangian}
\mathcal{L}_{\mathrm{ADM}}&=\mathcal{L}^{(0)}+\mathcal{L}^{(u)}+\mathcal{L}^{(s)}_1+\mathcal{L}_2^{(s)}+\mathcal{L}'^{(s)}\,, \displaybreak[0]\\[2ex]
\mathcal{L}^{(0)}&=\frac{N\sqrt{h}}{2}\bigg(\frac{2}{N}\mathcal{L}_mK-\frac{2}{N}D_iD^iN+R-2\Lambda \notag \\
&\phantom{{}={}}\hspace{1.1cm}+K^2+K_{ij}K^{ij}\bigg)\,, \displaybreak[0]\\[2ex]
\label{eq:theory-definition-u}
\mathcal{L}^{(u)}&=\frac{N\sqrt{h}}{2}\,\bigg[-u(R-K^2+K_{ij}K^{ij}) \notag \\
&\phantom{{}={}}\hspace{1.1cm}+\frac{2}{N}(K\mathcal{L}_mu+uD_iD^iN)\bigg]\,, \displaybreak[0]\\[2ex]
\label{eq:theory-definition-snn}
\mathcal{L}^{(s)}_1&=\frac{N\sqrt{h}}{2}\bigg[-\frac{1}{N}(K_{ij}\mathcal{L}_ms^{ij}+s^{ij}D_iD_jN) \notag \\
&\phantom{{}={}}\hspace{1cm}+s^{ij}(R_{ij}-2K_i^{\phantom{i}l}K_{lj})\bigg]\,, \displaybreak[0]\\[2ex]
\label{eq:theory-definition-sij}
\mathcal{L}^{(s)}_2&=\frac{N\sqrt{h}}{2}\bigg[s^{\mathbf{nn}}\left(\frac{1}{N}D_iD^iN-K^{ij}K_{ij}+K^2\right) \notag \\
&\phantom{{}={}}\hspace{1cm}+\frac{1}{N}K\mathcal{L}_ms^{\mathbf{nn}}\bigg]\,, \displaybreak[0]\\[2ex]
\mathcal{L}'^{(s)}&=\frac{N\sqrt{h}}{2}\left[2s^{i\mathbf{n}}(D_iK-D_lK^l_{\phantom{l}i})\right]\,,
\end{align}
\end{subequations}
with $h:=\det(h_{ij})$, the Ricci tensor $R_{ij}$ defined in $\Sigma_t$, and the corresponding Ricci scalar $R:= R^i_{\phantom{i}i}$. The extrinsic-curvature tensor $K_{ij}$ is given by
\begin{equation}
\label{eq:curvatureext}
K_{i j} = \frac{1}{2 N} (\dot{h}_{i j} - D_i N_j - D_j N_i)\,,
\end{equation}
where the dot denotes a time derivative. We also employ its trace $K:=K^i_{\phantom{i}i}$ as well as the Lie derivative \cite{Carroll:1997ar} of the extrinsic-curvature tensor with respect to the four-vector $m^{\mu}:= Nn^{\mu}$. The latter reads
\begin{equation}
\mathcal{L}_m K_{i j} = \dot{K}_{i j} -\mathcal{L}_N K_{i j}\,,
\end{equation}
with the Lie derivative $\mathcal{L}_N$ for the shift vector. Furthermore, $s^{ij}:=h^i_{\phantom{i}\mu}h^j_{\phantom{j}\nu}s^{\mu\nu}$ is the piece of $s^{\mu\nu}$ that is completely projected into $\Sigma_t$. We will be referring to it as the (purely spacelike) tensorial part. Also, $s^{i\mathbf{n}}:=h^i_{\phantom{i}\mu}n_{\nu}s^{\mu\nu}$ is understood as a vector-valued part and $s^{\mathbf{nn}}:=s^{\mu\nu}n_{\mu}n_{\nu}$ as the purely timelike contribution, which lives in the direction orthogonal to $\Sigma_t$. The coefficients $s^{i\mathbf{n}}$ and $s^{\mathbf{nn}}$ are taken as new, independent degrees of freedom. An important thing to bear in mind is that the coefficients $s^{i\mathbf{n}}$ were demonstrated to be gauge degrees of freedom~\cite{Reyes:2021cpx}. Therefore, $\mathcal{L}'^{(s)}$ of Eq.~\eqref{eq:theory-definition-sij} will be discarded in the remainder of the paper.

It is valuable to reformulate the action of Eq.~\eqref{eq:ADM-action} such that the Lie derivatives act on the background fields instead of the extrinsic curvature. To do so, the following identities are valuable:
\begin{subequations}
\label{eq:deruelle-relations}
\begin{align}
u \left( \frac{2}{N} \mathcal{L}_m K \right) &= 2 \nabla_{\mu} (n^{\mu} K u) -
2 u K^2 - \frac{2}{N} K\mathcal{L}_m u\,, \displaybreak[0]\\
s^{i j} \left( \frac{1}{N} \mathcal{L}_m K_{i j} \right) &= \nabla_{\mu}
(n^{\mu} K_{i j} s^{i j}) - K K_{i j} s^{i j} \notag \\
&\phantom{{}={}}- \frac{1}{N} K_{i j}
\mathcal{L}_m s^{i j}\,, \displaybreak[0]\\
- s^{\mathbf{nn}} \left( \frac{1}{N} \mathcal{L}_m K \right) &=
- \nabla_{\mu} (n^{\mu} K s^{\mathbf{nn}}) + K^2 s^{\mathbf{nn}} \notag \\
&\phantom{{}={}}+ \frac{1}{N} K\mathcal{L}_m s^{\mathbf{nn}}\,.
\end{align}
\end{subequations}
In this context, we must refer to another important property of Eq.~\eqref{eq:actionus}, which was emphasized in Ref.~\cite{Reyes:2021cpx}. By adding suitable modified Gibbons-Hawking-York boundary terms~\cite{York:1972sj,Gibbons:1976ue,Blau:2002,Heisenberg:2018vsk} to the ADM-decomposed action, the total derivatives in Eq.~\eqref{eq:deruelle-relations} can be discarded.

As a first step of our study, we introduce some simplifications to the modified-gravity theory defined by Eq.~\eqref{eq:actionus}. Let us impose the following conditions on the background fields:
\begin{equation}
\label{eq:consistency-conditions}
\mathcal{L}_m u =\mathcal{L}_m s^{i j} =\mathcal{L}_m s^{\mathbf{nn}} = 0\,.
\end{equation}
As we shall see later, the latter requirements imply that $u$ and $s^{\mu\nu}$ are time-independent in Gaussian normal coordinates \cite{Misner:1973}, which leads to vast computational simplifications. From a physical perspective, the background fields are then static and remain unmodified in the course of cosmological expansion. We think that this scenario is a reasonable point to start a first investigation from, but Eq.~\eqref{eq:consistency-conditions} will be dropped in a forthcoming chapter of the paper.

We also redefine the extrinsic curvature via
\begin{equation}
K_{i j} =: \frac{E_{i j}}{N}\,,\quad K =: \frac{E}{N}\,,
\end{equation}
with the quantities $E_{ij}$ and $E$ that are frequently employed in cosmology. As $K_{ij}$ and $K$ scale with $1/N$, the latter $E_{ij}$ and $E$ do not depend on the lapse function, anymore. Based on all these ingredients, the action of Eq.~\eqref{eq:ADM-action} can be cast into the form
\begin{subequations}
\label{eq:ADM-action-reformulated}
\begin{align}
S&= \int_{\mathcal{M}} \mathrm{d}t\mathrm{d}^3x\,(\mathcal{L}^{(0)}+\mathcal{L}^{(u)}+\mathcal{L}^{(1)}+\mathcal{L}^{(2)})+ S_m\,, \displaybreak[0]\\[2ex]
\mathcal{L}^{(0)}&=\frac{\sqrt{h}}{2}N\left[R-2\Lambda + \frac{1}{N^2} (E_{i j} E^{i j} - E^2) \right]\,, \displaybreak[0]\\[2ex]
\mathcal{L}^{(u)}&=\frac{\sqrt{h}}{2}N\bigg\{\frac{2}{N}\left(\frac{E}{N}\mathcal{L}_mu+uD_iD^iN\right) \notag \\
&\phantom{{}={}}\hspace{1.1cm}- u\left[ R + \frac{1}{N^2} (E_{i j} E^{i j} - E^2) \right]\bigg\}\,, \displaybreak[0]\\[2ex]
\mathcal{L}^{(s)}_1&=\frac{\sqrt{h}}{2}N\bigg[-\frac{1}{N}\left(\frac{E_{ij}}{N}\mathcal{L}_ms^{ij}+s^{ij}D_i D_j N\right) \notag \\
&\phantom{{}={}}\hspace{1.1cm}+s^{ij} \left(R_{ij} - \frac{2}{N^2} E_i^{\phantom{i}l} E_{l j}\right)\bigg]\,, \displaybreak[0]\\[2ex]
\mathcal{L}^{(s)}_2&=\frac{\sqrt{h}}{2}N\bigg[\frac{1}{N}\left(\frac{E}{N}\mathcal{L}_ms^{\mathbf{nn}}+s^{\mathbf{nn}}D_i D^i N\right) \notag \\
&\phantom{{}={}}\hspace{1.1cm}+\frac{s^{\mathbf{nn}}}{N^2}(E^2-E^{ij}E_{ij})\bigg]\,.
\end{align}
\end{subequations}
Matter ought to be modeled as a perfect fluid described by an energy-momentum tensor $(T_m)^{\mu\nu}$ that is chosen as~\cite{Misner:1973}
\begin{equation}
(T_m)^{\mu\nu}=(\rho+P)U^{\mu}U^{\nu} + P g^{\mu\nu}\,,
\end{equation}
where $\rho$ is the fluid density, $P$ its pressure, and $U^{\mu}$ its four-velocity. It is common to consider a fluid at rest such that $U^\mu=(1,0,0,0)$. Hence,
\begin{subequations}
\begin{align}
(T_m)^{00}&=\rho + (1 + g^{00})P\,, \\[1ex]
(T_m)^{0i}&=g^{0i}P\,, \\[1ex]
(T_m)^{ij}&=g^{ij}P\,.
\end{align}
\end{subequations}
In contrast to the gravity sector, the matter sector is conventional, whereupon we will take this choice, too. The stress-energy tensor is computed from the matter action $S_m$, as usual:
\begin{equation}
(T_m)_{\mu\nu}=-\frac{2}{\sqrt{-g}}\frac{\delta S_m}{\delta g^{\mu\nu}}\,.
\end{equation}
In the ADM formalism this relationship is equivalent to
\begin{equation}
\frac{\delta S_m}{\delta g^{\mu\nu}}=-\frac{\sqrt{h}}{2}N(T_m)_{\mu\nu}\,.
\end{equation}
By using $\delta g^{00}/\delta N=2/N^3$ and $\delta g^{0i}/\delta N^i=1/N^2$, the variations of the matter action with respect to the lapse function and the shift vector, respectively, read
\begin{subequations}
\label{eq:functional-derivatives-matter-action}
\begin{align}
-\frac{N^2}{\sqrt{h}}\frac{\delta S_m}{\delta N} &=(N^2-N_iN^i)^2\rho \notag \\
&\phantom{{}={}}+ \bigg[(N^2 - N_iN^i)^2 \left(1 - \frac{1}{N^2}\right) \notag \\
&\phantom{{}={}}\hspace{0.6cm}- 2 (N^2 - N_i N^i) \frac{N_i N^i}{N^2} \notag \\
&\phantom{{}={}}\hspace{0.6cm}+ N_iN_j \left(h^{ij} - \frac{N^iN^j}{N^2}\right)\bigg]P\,, \displaybreak[0]\\[2ex]
-\frac{2N}{\sqrt{h}}\frac{\delta S_m}{\delta N^i}&=(N_kN^k-N^2)N_i\rho \notag \displaybreak[0]\\
&\phantom{{}={}}+ \bigg[  (N_k N^k-N^2) N_i \left( 1 - \frac{1}{N^2}
\right) \notag \\
&\phantom{{}={}}\hspace{0.6cm}+ (N_k N^k-N^2) h_{i j} \frac{N^j}{N^2} + N_i N_j \frac{N^j}{N^2} \notag \displaybreak[0]\\
&\phantom{{}={}}\hspace{0.6cm}+ N_j h_{i k} \left( h^{j k} - \frac{N^j N^k}{N^2} \right) \bigg]P\,.
\end{align}
\end{subequations}
As of now, we will take the conditions of Eq.~\eqref{eq:consistency-conditions} into account. Then, a variation of Eq.~\eqref{eq:ADM-action-reformulated} for $N$ gives
\begin{subequations}
\begin{align}
\label{eq:constraint-1}
0 &= (1 - u) \left[ R - \frac{1}{N^2} (E_{i j} E^{i j} - E^2) \right] \notag \\
&\phantom{{}={}}+2 D_i D^i u - D_i D_j s^{i j} + s^{i j} R_{i j} \notag \\
&\phantom{{}={}}+\frac{2}{N^2} s^{i j} E_i^{\phantom{i}l}
E_{l j} + D_i D^i s^{\mathbf{nn}} \notag \\
&\phantom{{}={}}+ \frac{1}{N^2}s^{\mathbf{nn}} (E^{i j} E_{i j} - E^2) -2\Lambda+\frac{2}{\sqrt{h}}\frac{\delta S_m}{\delta N}\,,
\end{align}
and varying Eq.~\eqref{eq:ADM-action-reformulated} for $N_{k}$ implies
\begin{align}
\label{eq:constraint-2}
0&= 2 D_i \bigg\{ \frac{1}{N} \Big[(1 - u - s^{\mathbf{nn}}) (E^{i
k}_{} - h^{i k} E) \notag \\
&\phantom{{}={}}\hspace{0.8cm}- s^{i j} E^k_{\phantom{a}j} - s^{j k} E^i_{\phantom{i}j}\Big] \bigg\}+\frac{2}{\sqrt{h}}\frac{\delta S_m}{\delta N^k}\,.
\end{align}
\end{subequations}
The latter constraints will play a significant role in the forthcoming phenomenological analysis.

\section{Cosmology with diffeomorphism-violating background fields}\label{sec:cosmo}

To study the implications of the modified-gravity theory of Eq.~\eqref{eq:actionus} on cosmology, we would have to solve the associated modified Einstein equations, which is a highly challenging task. To avoid this arduous pathway, we will be working with a Friedmann-Lema\^{i}tre-Robertson-Walker (FLRW) metric without perturbations:
\begin{subequations}
\label{eq:FLRWline}
\begin{equation}
\mathrm{d}s^2=-\mathrm{d}t^2 + h_{ij}\mathrm{d}x^i\mathrm{d}x^j\,,\quad h_{ij}=a(t)^2\tilde{g}_{ij}\,,
\end{equation}
with the cosmic scale factor $a(t)$ and the time-independent spatial part of the spacetime metric given by
\begin{equation}
\label{eq:FLRW-spatial}
\tilde{g}_{ij}=\left(\frac{\mathrm{d}r^2}{1 - k r^2} + r^2\,\mathrm{d}\theta^2 +
r^2 \sin^2 \theta\,\mathrm{d}\phi^2\right)\,,
\end{equation}
\end{subequations}
in three-dimensional spherical coordinates $(r,\theta,\phi)$ where $k\in \{-1,0,1\}$ represents the scalar curvature of an open, flat, and closed Universe, respectively. The FLRW metric rests on the assumptions of homogeneity of spacetime and spatial isotropy. Thus, at first it seems odd why the FLRW metric should be compatible with Eq.~\eqref{eq:actionus}, which violates diffeomorphism invariance as well as local Lorentz invariance in freely falling inertial reference frames.

However, one must keep in mind that deviations from GR have not been found, so far. The background fields $u$ and $s^{\mu\nu}$ are effective descriptions of possible Planck scale phenomena. So it is a very reasonable assumption that alterations of the FLRW metric of Eq.~\eqref{eq:FLRWline} are strongly suppressed. Then the FLRW metric is a wise choice to base a first study of the cosmological effects of Eq.~\eqref{eq:actionus} on. This procedure is corroborated by the tight experimental bounds on SME coefficients compiled in Ref.~\cite{Kostelecky:2008bfz}.

By employing the form of Eq.~\eqref{eq:FLRWline}, a direct comparison between Eq.~\eqref{eq:admline} and Eq.~\eqref{eq:FLRWline} reveals that $N = 1$ and $N^i = 0$. These choices correspond to using Gaussian normal coordinates. Then, the functional derivatives of the ADM-decomposed matter action presented in Eq.~\eqref{eq:functional-derivatives-matter-action} collapse to more convenient results:
\begin{subequations}
\begin{align}
\frac{\delta S_m}{\delta N} &= -\sqrt{h} \rho\,, \\[1ex]
\frac{\delta S_m}{\delta N^i}&= 0\,.
\end{align}
\end{subequations}
Furthermore, a straightforward computation of the Ricci tensor and Ricci scalar of the spatial part of the FLRW metric stated in Eq.~\eqref{eq:FLRW-spatial} results in
\begin{subequations}
\begin{align}
R_{ij}&=\frac{2k}{a^2}h_{ij}\,, \\[1ex]
R&=6\frac{k}{a^2}\,.
\end{align}
\end{subequations}
A slew of valuable identities follows from Eq.~\eqref{eq:curvatureext} and the metric of Eq.~\eqref{eq:FLRWline}:
\begin{subequations}
\label{eq:formulas-extrinsic-curvature}
\begin{align}
E_{i j}&=H h_{i j}\,, \displaybreak[0]\\[2ex]
\dot{E}_{i j} &= 2 a \dot{a} H \tilde{g}_{i j} + a^2 \dot{H} \tilde{g}_{i j} = (2 H^2 + \dot{H})h_{i j}\,, \displaybreak[0]\\[2ex]
\dot{E}^{ij}&=\frac{\mathrm{d}}{\mathrm{d}t}(E_{kl}h^{ki}h^{lj})=\dot{E}_{kl}h^{ki}h^{lj}+E^i_{\phantom{i}l}\dot{h}^{lj}+E_k^{\phantom{k}j}\dot{h}^{ki} \notag \\
&=(2 H^2 + \dot{H})h^{ij}-4H^2h^{ij} \notag \\
&=(-2H^2+\dot{H})h^{ij}\,, \displaybreak[0]\\[2ex]
\dot{E}&=\frac{\mathrm{d}}{\mathrm{d}t}(E_{ij}h^{ij})=\dot{E}_{ij}h^{ij}+E_{ij}\dot{h}^{ij} \notag \\
&=3(2H^2+\dot{H})-6H^2=3\dot{H}\,, \displaybreak[0]\\[2ex]
E_{i j} E^{i j} &= h^{r i} h^{s j} E_{i j} E_{r s} \notag \\
&= \frac{1}{a^4} \delta^{r
i} \delta^{s j} a^4 H^2 \delta_{i j} \delta_{r s}= 3 H^2\,, \displaybreak[0]\\[2ex]
E &= h^{i j} E_{i j} = \frac{1}{a^2} \delta^{i j} a^2 H \delta_{i j} = 3 H\,, \displaybreak[0]\\[2ex]
E_i^{\phantom{i}l} E_{l j} &= h^{l r} E_{i r} E_{l j} = H^2 h_{ij}\,,
\end{align}
\end{subequations}
where $H:=\dot{a}/a$ is the Hubble parameter. These relations allow us to reformulate the constraints of Eqs.~(\ref{eq:constraint-1}), (\ref{eq:constraint-2}). Doing so for the first leads to
\begin{subequations}
\label{eq:constraint1}
\begin{align}
H^2&=\frac{1}{3\Xi}\left(\rho+\Lambda-3\Upsilon\frac{k}{a^2}-D_iD^iu \right. \notag \\
&\phantom{{}={}}\hspace{0.8cm}\left.{}-\frac{1}{2}D_iD^is^{\mathbf{nn}}+\frac{1}{2}D_iD_js^{ij}\right)\,,
\end{align}
with the definitions
\begin{equation}
\Upsilon:= 1 - u + \frac{s}{3}\,,\quad \Xi:= \Upsilon-s^{\mathbf{nn}}\,,
\end{equation}
\end{subequations}
which are introduced for brevity. Furthermore, we define the trace of $s^{ij}$ via $s:=s^{i j} h_{i j} = a^2 s^{i j} \tilde{g}_{i j}$. The reformulated second constraint reads
\begin{equation}
\label{eq:constraint2}
0 = 4 D_i \bigg\{H \bigg[(1 - u - s^{\mathbf{nn}}) h^{i k} + s^{i k}\bigg]\bigg\}\,.
\end{equation}
Equation~\eqref{eq:constraint1} is interpreted as the first modified Friedmann equation of a cosmological evolution based on Eq.~\eqref{eq:actionus}. Notice that the first standard Friedmann equation with cosmological constant is recovered when we set the background fields $u$, $s^{i j}$, and $s^{\mathbf{n}\mathbf{n}}$ to zero. Note that Eq.~\eqref{eq:constraint2} is automatically satisfied in the standard case, but in the presence of background fields it poses an additional constraint that must be taken into account.

We highlight that spacetime homogeneity and spatial isotropy of GR imply a diagonal energy-momentum tensor. As a consequence, energy-momentum conservation for matter, $\nabla_\mu (T_m)^{\mu\nu}=0$, leads to
\begin{equation}
\label{eq:continuity}
\dot{\rho} = - 3 H (\rho + P)\,.
\end{equation}
With the standard equation of state $P = w \rho$ where $w$ is a characteristic parameter for matter, radiation, etc., Eq.~\eqref{eq:continuity} allows us to deduce a relationship between the matter density and the scale factor:
\begin{equation}
\label{eq:rhowitha}
\rho =\frac{\rho_0(a_0)^{3 (1 + w)}}{a^{3 (1 + w)}}\,,
\end{equation}
where $a_0=a(t_0)$ and $\rho=\rho(t_0)$ at an initial time $t_0$.
In the modified setting based on Eq.~(\ref{eq:actionus}), homogeneity is lost, as the background fields in a curved spacetime manifold necessarily depend on the coordinates \cite{Kostelecky:2003fs}. This property implies diffeomorphism violation, after all. As mentioned before, if a background vierbein is employed to define background fields from $s^{\mu\nu}$ in freely falling inertial frames \cite{Kostelecky:2020hbb}, even spatial isotropy is lost in most cases. Thus, a constraint of the form of Eq.~(\ref{eq:continuity}) does not necessarily apply to the modified scenario, anymore. So we are free to choose a different class of energy-momentum tensor. However, we will continue describing matter as a perfect fluid, as usual, since matter is assumed to be standard and the gravity-sector background fields supposedly involve minuscule component coefficients. Indeed, a description of matter via a perfect fluid is also convenient from a technical point of view, which is why Eq.~\eqref{eq:continuity} shall be taken over to our setting.

Now, we turn towards deriving the second modified Friedmann equation. The second Friedmann equation in GR is a suitable linear combination of the first Friedmann equation and the dynamical part of the Einstein equations. The dynamics of GR is encoded in the purely spacelike part of the Einstein equations, i.e.,
\begin{equation}
\label{eq:purely-spacelike-einstein-equations}
G^{ij}+\Lambda g^{ij}=(T_m)^{ij}\,,\quad G^{ij}=R^{ij}-\frac{R}{2}g^{ij}\,,
\end{equation}
where $G^{ij}$ is the spatial part of the Einstein tensor. Alternatively, to understand the dynamics of Einstein's gravity, the Hamilton equation
\begin{equation}
\dot{\pi}^{ij}=\{\pi^{ij},H\}\,,
\end{equation}
can be consulted where $H$ is the GR Hamiltonian based on the ADM decomposition and $\pi^{ij}$ is the canonical momentum density \cite{Bertschinger:2005}. Here, $\{A,B\}$ denote suitably defined Poisson brackets of the (tensor-valued) objects $A$ and $B$, which are functions of $h_{kl},\pi^{mn}$, etc. and whose index structures are omitted for brevity. Both approaches provide the same dynamical equations.

An intriguing property of GR is that the second Friedmann equation can be shown to be a consequence of the first Friedmann equation when energy-momentum conservation for matter, i.e., Eq.~\eqref{eq:continuity}, is employed. In particular, differentiating the first Friedmann equation for time, using energy-momentum conservation, and inserting the first Friedmann equation again subsequently gives rise to the second Friedmann equation \cite{Misner:1973}. Therefore, the second Friedmann equation in GR can be regarded as superfluous and the focus is usually on the first Friedmann equation only.

Diffeomorphism invariance is a crucial characteristic of GR and the diffeomorphism group provides a considerable symmetry structure that many of the interesting properties of GR are based on. What holds in GR is not necessarily valid when diffeomorphism symmetry breaks down. For example, the constraint structure and number of physical degrees of freedom is subject to significant alterations, as demonstrated, e.g., in $f(Q)$ gravity~\cite{Hu:2022anq}. In our context, as it turns out, the dynamical part of the modified Einstein equations of Eq.~\ref{eq:actionus} in combination with Eq.~\eqref{eq:constraint1} provide a result that differs from the equation that follows from differentiating Eq.~\eqref{eq:constraint1} for time and using energy-momentum conservation. So we interpret the remarkable property of the second Friedmann equation in GR being an implication of the first Friedmann equation plus energy-momentum conservation as a consequence of diffeomorphism invariance.

As it is the modified Einstein equations that are fundamental, we employ the latter to derive the second modified Friedmann equations of Eq.~\eqref{eq:actionus} without resorting to the time derivative of Eq.~\eqref{eq:constraint1}. The recent results of Ref.~\cite{Reyes:2022mvm} are highly beneficial to perform these calculations based on the FLRW metric. We will also need to evaluate various Lie derivatives of background field coefficients. Notice that the Lie derivative of a scalar field amounts to a directional derivative as follows:
\begin{align}
\mathcal{L}_m u&= m^{\mu} \partial_{\mu} u = N n^{\mu}\partial_{\mu} u \notag \\
&= N n^0 \dot{u} + N n^i D_i u = \dot{u} - N^i D_i u\,.
\end{align}
Here we have used that $N n^0=1$ and $N n^i=-N^i$. Furthermore, as $N=1$ and $N^i = 0$ hold in a nonperturbed FLRW metric, we deduce $\mathcal{L}_m=\dot{u}$ directly. Now, turning to $s^{\mu\nu}$, we consult Ref.~\cite{Carroll:1997ar} for the definition of the Lie derivative for a rank-2 tensor-valued background field:
\begin{align}
\mathcal{L}_m s^{\rho \sigma} &= m^{\mu} \partial_{\mu} s^{\rho \sigma} -
(\partial_{\lambda} m^{\rho}) s^{\lambda \sigma} - (\partial_{\lambda}
m^{\sigma}) s^{\rho \lambda} \notag \\
&= \dot{s}^{\rho \sigma} - N^i \partial_i s^{\rho
\sigma} - (\partial_{\lambda} m^{\rho}) s^{\lambda \sigma} -
(\partial_{\lambda} m^{\sigma}) s^{\rho \lambda}\,.
\end{align}
But $\partial_{\lambda} m^{\rho} = 0$, since the component $m^0$ is constant and the components $m^i$ depend on $N^i$, which are zero in a nonperturbed FLRW metric. Thus, the Lie derivative amounts to a simple time derivative: $\mathcal{L}_m s^{\rho \sigma} =\dot{s}^{\rho \sigma}$.

So in a FLRW spacetime, Eq.~\eqref{eq:consistency-conditions} translates to $\dot{u}=\dot{s}^{\mathbf{nn}}=\dot{s}^{ij} = 0$. Then, each background field is taken as static. In addition, we deduce the following valuable relation for the time derivative of the trace of~$s^{ij}$:
\begin{equation}
\label{eq:time-derivative-trace}
\dot{s} = 2 a \dot{a} s^{i j}\tilde{g}_{i j} + a^2 \dot{s}^{i j} \tilde{g}_{i j} = 2 a^2 H s^{i j}\tilde{g}_{i j} = 2 H s\,.
\end{equation}
As the trace involves the FLRW metric by definition, its time derivative does not simply vanish, but it is proportional to the trace itself. Another useful relationship is $\ddot{s}=2s(\dot{H}+2H^2)$, which expresses the second-order time derivative of the trace $s$ in terms of the trace proper.

The modified Einstein equations for the action of Eq.~\eqref{eq:actionus} with $\Lambda=0$ are stated in Ref.~\cite{Bailey:2006fd}. Their purely spacelike part follows from the findings of Sec.~IV in Ref.~\cite{Reyes:2022mvm} with Eq.~\eqref{eq:formulas-extrinsic-curvature} and the latter results for the Lie derivatives of the controlling coefficients taken into account. The first Friedmann equation in Eq.~\eqref{eq:constraint1} is then divided by 2 and subtracted from the dynamical equations leading to a perturbation of the second Friedmann equation of GR. For background fields satisfying Eq.~\eqref{eq:consistency-conditions}, we obtain
\begin{align}
\label{eq:secfriedmannb}
\dot{H}+H^2&=-\frac{1}{6\Xi}\left[\rho+3P-2\Lambda+2s\left(\frac{k}{a^2}+H^2\right)\right. \notag \\
&\phantom{{}={}}\hspace{0.9cm}\left.{}+D_iD^iu-\frac{1}{2}D_iD^is^{\mathbf{nn}}-\frac{1}{2}D_iD^is\right]\,,
\end{align}
where $\dot{H} + H^2=\ddot{a}/a$. To the best of our knowledge, the set of modified Friedmann equations in Eqs.~\eqref{eq:constraint1}, \eqref{eq:secfriedmannb} based on the action of Eq.~\eqref{eq:actionus} has been derived here for the first time. In conjunction with the additional constraint of Eq.~\eqref{eq:constraint2} they are necessary to draw conclusions on the time evolution of a Universe whose gravitational laws are based on the action of Eq.~\eqref{eq:actionus}. For vanishing background fields, the second Friedmann equation with cosmological constant is reproduced, as expected.

Both equations involve the background fields as well as second-order spatial derivatives of the latter, but first-order derivatives do not occur. Also, when multiplying both Friedmann equations with $\Xi$, they are linear in the background fields, which is an implication from the action as well as the modified Einstein equations being linear in the coefficients $u,s^{\mathbf{nn}},s^{ij}$. For completeness and reasons of comparison, the field equations that follow from differentiating Eq.~\eqref{eq:constraint1} for time as well as using energy-momentum conservation for matter will be stated in App.~\ref{sec:alternative-equations}.

\section{Study of accelerated expansion of the Universe}\label{sec:inflation}

In what follows, we will particularly be interested in understanding whether or not the background fields $u$ and $s^{\mu\nu}$ of Eq.~\eqref{eq:actionus} are capable of driving an accelerated expansion of a Universe that contains ordinary matter and radiation only. Thus, we will set $\Lambda=0$. The inflationary regime immediately after the Big Bang as well as observations of the current state of our Universe hint towards the existence of accelerated stages~\cite{SupernovaSearchTeam:1998fmf,SupernovaCosmologyProject:1998vns,SupernovaCosmologyProject:2008ojh,Pan-STARRS1:2017jku,WMAP:2003elm,WMAP:2008lyn,Planck:2018vyg,SDSS:2005xqv,eBOSS:2020yzd,DES:2021wwk}. Mathematically, the condition of an accelerated expansion can be described in terms of the scale factor as
\begin{equation}
\label{eq:acexp}
\ddot{a} > 0\,,
\end{equation}
which translates into a negative deceleration parameter:
\begin{equation}
\label{eq:deceleration-parameter}
q:= -\frac{\ddot{a} a}{\dot{a}^2}<0\,.
\end{equation}
In the cosmological standard model, by taking the second Friedmann equation and the condition of Eq.~\eqref{eq:acexp} into consideration, we get the following inequality between the pressure $P$ and the energy density $\rho$ of matter:
\begin{equation}
\label{eq:secbreak}
P < - \frac{\rho}{3}\,.
\end{equation}
With this finding at hand, we infer that a stage of accelerated expansion in standard cosmology is caused by an entity that breaks the Strong Energy Condition (SEC)~\cite{Visser:1999de}, which implies it having a negative pressure. Since standard matter respects the SEC, the usual way to proceed in inflation is to add a scalar field whose potential dominates over the kinetic term in a slow-roll regime~\cite{Baumann:2009ds,Liddle:2009}. The common way to implement an accelerated stage into a particular model of our Universe is to resort to Dark Energy. Incorporating a cosmological constant into the action, this source can be described by means of an equation of state $P=-\rho$ satisfying the inequality of Eq.~\eqref{eq:secbreak}. On the other hand, different sources can be used to mimic the effects of Dark Energy such as new scalar fields or fluids with different equations of state (see the reviews \cite{Bamba:2012cp,Copeland:2006wr} for details). However, in any case, a new kind of matter that violates the SEC is introduced.

In the context of the gravitational SME, we have found that there are changes in the Friedmann equations (see Eqs.~\eqref{eq:constraint1} and \eqref{eq:secfriedmannb}) that have the potential to drive a quite different scenario compared to a setting with only standard matter and radiation. We will analyze separate cases for the sake of simplicity. Besides, in accordance with experimental measurements of the curvature of our Universe \cite{SDSS:2005xqv,WMAP:2008lyn,Planck:2018vyg}, $k=0$ is to be employed in our forthcoming analyses.

\subsection{Scalar background $u$}\label{sec:utime}

First of all, we consider a scenario where $s^{\mu\nu}=0$ such that we are able to focus on the scalar background field~$u$. The modified Einstein equations for this case are \cite{Bailey:2006fd}
\begin{subequations}
\begin{equation}
{}^{(4)}G^{\alpha \beta} = (T_m)^{\alpha \beta} + (T^{Ru})^{\alpha \beta}\,,
\end{equation}
with the four-dimensional Einstein tensor ${}^{(4)}G^{\alpha \beta}$, the energy-momentum tensor $(T_m)^{\alpha \beta}$ for matter, and the characteristic two-tensor
\begin{equation}
(T^{Ru})^{\alpha \beta} = - \nabla^{\alpha} \nabla^{\beta} u + g^{\alpha \beta}\nabla^2 u + u {}^{(4)}G^{\alpha \beta}\,,
\end{equation}
\end{subequations}
that involves the background field. Computing the covariant derivative of the latter leads to
\begin{align}
\nabla_{\alpha} (T^{Ru})^{\alpha \beta} &= - \nabla^2 \nabla^{\beta} u +
(\nabla_{\alpha} g^{\alpha \beta}) \nabla^2 u + \nabla^{\beta} \nabla^2 u \notag \\
&\phantom{{}={}}+(\nabla_{\alpha} u){}^{(4)}G^{\alpha \beta}\,,
\end{align}
whereby metric compatibility implies
\begin{equation}
\nabla_{\alpha} (T^{Ru})^{\alpha \beta} = (\nabla_{\alpha} u) {}^{(4)}G^{\alpha\beta}\,.
\end{equation}
At this point we are coming to an essential feature of gravity theories modified by nondynamical background fields coupling to spacetime curvature. Recall the second Bianchi identities of (pseudo)-Riemannian geometry, which are frequently written as $\nabla_{\mu}{}^{(4)}G^{\mu\nu}=0$ in the context of GR. It is these identities that severely restrict the freedom of choosing a background field violating diffeomorphism invariance explicitly. Applying them to the modified Einstein equations provides a set of additional restrictions on the background field that are often denoted as no-go results in the literature \cite{Kostelecky:2003fs,Kostelecky:2020hbb,Bluhm:2014oua,Bluhm:2016dzm}. Because of these limitations as well as energy-momentum conservation in the matter sector, we have that
\begin{equation}
\label{eq:no-go-constraints-u}
\nabla_{\alpha} (T^{Ru})^{\alpha \beta} = 0 \Rightarrow
\nabla_{\alpha} u = 0\,,
\end{equation}
which provides $\nabla_iu=\partial_iu=0$ in addition to $\dot{u}=0$ already being satisfied due to the conditions of Eq.~(\ref{eq:consistency-conditions}). Then, the scalar field $u$ also obeys the additional constraint of Eq.~\eqref{eq:constraint2}.

Now, let us draw conclusions on a possible accelerated stage of the Universe from the modified Friedmann equations.
After using Eq.~\eqref{eq:secfriedmannb}, the condition of Eq.~\eqref{eq:acexp} reads
\begin{equation}
\frac{\ddot{a}}{a} = - \frac{1}{6 (1 - u)} (\rho + 3 P + D_i D^i u) >0\,.
\end{equation}
As long as gravitational fields are weak enough, we expect that $|u|\ll 1$ such that $1 - u > 0$. Hence, due to Eq.~\eqref{eq:no-go-constraints-u},
\begin{equation}
\rho + 3 P + D_i D^i u=\rho+3P < 0\,.
\end{equation}
Then, the Friedmann equations for $k=0$ read
\begin{subequations}
\begin{align}
H^2 &= \frac{1}{3 (1 - u)} \rho\,, \\[2ex]
\frac{\ddot{a}}{a} &= - \frac{1}{6 (1 - u)} (\rho + 3 P)\,.
\end{align}
\end{subequations}
In this case, defining an effective density and pressure via $\rho_{\mathrm{eff}}:=\rho/(1-u)$ and $P_{\mathrm{eff}}:=P/(1-u)$, respectively, cosmological expansion occurs in the same way as in the standard case without diffeomorphism violation described by~$u$. The latter background field then simply gives rise to a constant scaling factor that does not have any impact on the time evolution of the Universe.

\subsection{Purely timelike background $s^{\mathbf{nn}}$}\label{sec:snntime}

Now, we consider the sector where $u=0$ and $s^{ij}=0$, as well. Note that $s^{\mathbf{nn}}=s^{00}$ in Gaussian normal coordinates, but we will keep $s^{\mathbf{nn}}$ for notational consistency. In general, the modified Einstein equations for $s^{\mu\nu}$ read~\cite{Bailey:2006fd}
\begin{subequations}
\label{eq:modified-einstein-equation-s}
\begin{equation}
{}^{(4)}G^{\alpha \beta} = (T_m)^{\alpha \beta} + (T^{Rs})^{\alpha \beta}\,,
\end{equation}
with
\begin{align}
\label{eq:emts}
(T^{Rs})^{\alpha \beta} &= \frac{1}{2} (g^{\alpha \beta} s^{\mu \nu} R_{\mu \nu} +
\nabla_{\nu} \nabla^{\alpha} s^{\nu \beta} + \nabla_{\nu} \nabla^{\beta}
s^{\nu \alpha} \notag \\
&\phantom{{}={}}\hspace{0.3cm}- \nabla^2 s^{\alpha \beta} - g^{\alpha \beta} \nabla_{\mu} \nabla_{\nu} s^{\mu \nu})\,.
\end{align}
\end{subequations}
Such as for the $u$ sector discussed previously, no-go restrictions also play a role for the $s^{\mu\nu}$ sector. By taking the divergence of Eq.~\eqref{eq:emts}, we deduce
\begin{align}
\label{eq:dtabs}
\nabla_{\alpha} (T^{Rs})^{\alpha \beta} &= \frac{1}{2} [g^{\alpha \beta}
\nabla_{\alpha} (s^{\mu \nu} R_{\mu \nu}) + \nabla_{\alpha} \nabla_{\nu}
\nabla^{\alpha} s^{\nu \beta} \notag \\
&\phantom{{}={}}\hspace{0.3cm}+ \nabla_{\alpha} \nabla_{\nu} \nabla^{\beta}
s^{\nu \alpha} - \nabla_{\alpha} \nabla^2 s^{\alpha \beta} \notag \\
&\phantom{{}={}}\hspace{0.3cm}- g^{\alpha \beta} \nabla_{\alpha} \nabla_{\mu} \nabla_{\nu} s^{\mu \nu}]\,.
\end{align}
Evaluating the components of the latter for the FLRW spacetime of Eq.~\eqref{eq:FLRWline} explicitly, gives rise to
\begin{subequations}
\begin{align}
\nabla_{\alpha} (T^{Rs})^{\alpha 0} &= \frac{3 s^{\mathbf{nn}}}{a^2} \left( 2 \dot{a}\ddot{a} + a \dddot{a} \right)\,, \label{eq:ts1} \displaybreak[0]\\[2ex]
\nabla_{\alpha} (T^{Rs})^{\alpha 1} &= - \frac{3 \ddot{a}}{2 a^3} \partial_r s^{\mathbf{nn}}\,, \label{eq:ts2}\displaybreak[0]\\[2ex]
\nabla_{\alpha} (T^{Rs})^{\alpha 2} &= - \frac{3 \ddot{a}}{2 a^3 r^2}\partial_{\theta} s^{\mathbf{nn}}\,, \label{eq:ts3}\displaybreak[0]\\[2ex]
\nabla_{\alpha} (T^{Rs})^{\alpha 3} &= - \frac{3 \ddot{a}}{2 a^3 r^2 \sin^2 \theta}
\partial_{\phi} s^{\mathbf{nn}}\,. \label{eq:ts4}
\end{align}
\end{subequations}
The second Bianchi identities and energy-momentum conservation encoded in Eq.~\eqref{eq:continuity} imply
\begin{equation}
\label{eq:no-go-constraint-einstein-equation-s}
\nabla_{\alpha} (T^{Rs})^{\alpha \beta} = 0\,.
\end{equation}
Therefore, the only possibility of satisfying Eq.~\eqref{eq:constraint2} and Eq.~\eqref{eq:no-go-constraint-einstein-equation-s} for $\beta=1,2,3$ is $s^{\mathbf{nn}} = c_1$ with a constant $c_1$.
In addition, Eq.~\eqref{eq:ts1} equated to zero implies an additional condition that involves the FLRW scale factor:
\begin{equation}
2 \dot{a} \ddot{a} + a \frac{\mathrm{d}^3 a}{\mathrm{d} t^3} = 0 \Leftrightarrow \frac{\mathrm{d}}{\mathrm{d}t}\left( \ddot{a} a + \frac{1}{2} \dot{a}^2 \right) = 0\,.
\end{equation}
Therefore, we deduce that
\begin{equation}
\label{eq:fixa}
\ddot{a} a + \frac{1}{2} \dot{a}^2 = c_2\,,
\end{equation}
with another constant $c_2$. Equation \eqref{eq:fixa} establishes a requirement on the scale factor if $c_1 \neq 0$. Standard cosmology is provided by $c_1=0$,  which gets rid of the background field. As before, it is reasonable to assume $|s^{\mathbf{nn}}|\ll 1$ in the presence of weak gravitational fields. Consequently, it holds that $1 - s^{\mathbf{nn}} > 0$.
In this case, both Friedmann equations for $k=0$ are
\begin{subequations}
\begin{align}
\label{eq:resulting-friedmann-equations-s00-first}
H^2 &= \frac{1}{3 (1 - s^{\mathbf{nn}})} \left( \rho -
\frac{1}{2} D_i D^i s^{\mathbf{nn}} \right)\,, \\[2ex]
\label{eq:resulting-friedmann-equations-s00-second}
\dot{H} + H^2 &= - \frac{1}{6 (1 - s^{\mathbf{nn}})} \left(\rho + 3 P -\frac{1}{2} D_i D^i s^{\mathbf{nn}}\right)\,.
\end{align}
\end{subequations}
The constraint of Eq.~\eqref{eq:constraint2} as well as Eqs.~\eqref{eq:ts2} -- \eqref{eq:ts4} set to zero imply $D_is^{\mathbf{nn}}=\partial_is^{\mathbf{nn}}=0$. Then, by subtracting Eq.~\eqref{eq:resulting-friedmann-equations-s00-first} from Eq.~\eqref{eq:resulting-friedmann-equations-s00-second}, we arrive at
\begin{equation}
\label{eq:time-derivative-hubble-parameter}
\dot{H} = - \frac{\rho + P}{2 (1 - s^{\mathbf{nn}})}\,.
\end{equation}
Thus, the double covariant derivative of $s^{\mathbf{nn}}$ is completely eliminated from the modified Friedmann equations. We then observe that the term $1-s^{\mathbf{nn}}$ acts like a rescaling of the conventional time evolution. Therefore, a regime of accelerated expansion requires a suitable source, because baryonic matter and electromagnetic radiation respect the SEC.

Although this case resembles the standard scenario without any background fields present, $s^{\mathbf{nn}}$ cannot simply be absorbed into the matter density and pressure, which distinguishes $s^{\mathbf{nn}}$ from the background field $u$ of Sec.~\ref{sec:utime}. Note that Eq.~\eqref{eq:fixa} corresponds to an extra condition, which does not occur for $u$. The latter combined with Eq.~\eqref{eq:time-derivative-hubble-parameter} modifies the behavior of the scale factor. To see this, we use a standard equation of state for matter, $P = w \rho$, and write Eq.~\eqref{eq:time-derivative-hubble-parameter} in terms of the scale factor to obtain
\begin{equation}
\frac{\ddot{a}}{a} - \frac{\dot{a}^2}{a^2} = - \frac{1 + w}{2 (1 - s^{\mathbf{nn}})} \rho\,.
\end{equation}
Inserting Eq.~\eqref{eq:fixa} results in
\begin{equation}
\label{eq:time-derivative-hubble-parameter-additional-condition}
\frac{1}{a^2} \left( c_2 - \frac{3 \dot{a}^2}{2} \right) = - \frac{1 + w}{2
(1 - s^{\mathbf{nn}})} \rho\,.
\end{equation}
It is now helpful to employ the previously derived Eq.~\eqref{eq:rhowitha}, which is an implication of energy-momentum conservation, in Eq.~\eqref{eq:time-derivative-hubble-parameter-additional-condition}:
\begin{equation}
\label{eq:time-derivative-scale-factor-snn}
\dot{a}^2 = c_3a^{- (1 + 3 w)} + \frac{2}{3}c_2\,,\quad c_3=\frac{1 + w}{3 (1 - s^{\mathbf{nn}})}
\rho_0(a_0)^{3 (1 + w)}\,.
\end{equation}
A solution of this differential equation with negative deceleration parameter of Eq.~\eqref{eq:deceleration-parameter} was not found. For a Universe without matter present, meaning $\rho=0$, we have $c_3=0$. Then, the first contribution in Eq.~\eqref{eq:time-derivative-scale-factor-snn} is eliminated and we arrive at
\begin{equation}
\dot{a}^2 = \frac{2}{3} c_2\,.
\end{equation}
A necessary condition for a physical solution is $c_2\geq 0$, whereupon $a \propto t$ if $c_2\neq 0$. This scale factor describes an empty Universe --- with only the background field $s^{\mathbf{nn}}$ present --- that expands linearly. A behavior of this kind resembles that of a Milne Universe \cite{Milne:1935}.

\subsection{Tensor-valued purely spacelike background $s^{ij}$}
\label{sec:sijb}

Finally, we consider a scenario with $u=s^{\mathbf{nn}}=0$. Since $s^{ij}$ is described by 6 independent coefficients, a general study of its impact is cumbersome. To gain an initial understanding of its implications, we have to respect the additional constraint of Eq.~\eqref{eq:constraint2} that requires
\begin{equation}
\label{eq:constraintsij}
D_i s^{ij}=0\,.
\end{equation}
Rearranging Eq.~\eqref{eq:secfriedmannb} with Eq.~(\ref{eq:constraintsij}) taken into account as well as $k=0$ inserted leads to
\begin{subequations}
\label{eq:friedmann-spatial}
\begin{align}
\label{eq:friedmann-1-spatial}
H^2 &= \frac{\rho}{3+s}\,, \\[2ex]
\dot{H}+H^2&=-\frac{1}{6(1+s/3)}\left(\rho+3P-\frac{1}{2}D_iD^is+2sH^2\right)\,.
\end{align}
\end{subequations}
A stage of accelerated expansion now emerges when either the first or the second of the following sets of conditions is satisfied:
\begin{subequations}
\begin{align}
\rho + 3 P &\lessgtr \frac{1}{2} D_i D^i s - 2 s H^2\,, \\[2ex]
\label{eq:bound-s}
s &\gtrless -3\,.
\end{align}
\end{subequations}
Inserting the first Friedmann equation and using the equation of state $P=w\rho$ for standard matter and radiation leads to a chain of inequalities:
\begin{align}
\label{eq:inequalities}
\frac{1}{2} D_i D^i s - \frac{2 s}{3+s} \rho &< 0 < (1 + 3 w) \rho \notag \\
&< \frac{1}{2} D_i D^i s - \frac{2 s}{3+s} \rho\,.
\end{align}
Let us now choose a particular example for a background field $s^{\mu\nu}$ with nonzero purely spacelike entries only:
\begin{equation}
\label{eq:sijnew}
s^{\mu \nu} = -\alpha \left(\begin{array}{cccc}
  0 & 0 & 0 & 0 \\
  0 & 1 & 0 & 0 \\
  0 & 0 & 1/r^2 & 0 \\
  0 & 0 & 0 & 1/(r^2 \sin^2 \theta) \\
\end{array}\right)\,,
\end{equation}
where $\alpha$ is a dimensionless real parameter. The latter tensor satisfies Eq.~\eqref{eq:constraintsij} and $s=-3\alpha a^2$. Using this value in Eq.~\eqref{eq:bound-s},
the condition for the scale factor describing an accelerated expansion of the Universe translates to
\begin{equation}
\label{eq:infws2}
a^2<\frac{1}{\alpha}\,,
\end{equation}
such that for $\alpha$ fixed, the scale factor must not exceed a certain value to enable accelerated expansion.
The trace $s$ in Eq.~\eqref{eq:bound-s} taking rather large (negative) values does not contradict $|s^{ij}|\ll 1$, as the size of $s^{ij}$ is controlled by $\alpha$.
\begin{figure}
\centering
\includegraphics[scale=0.4]{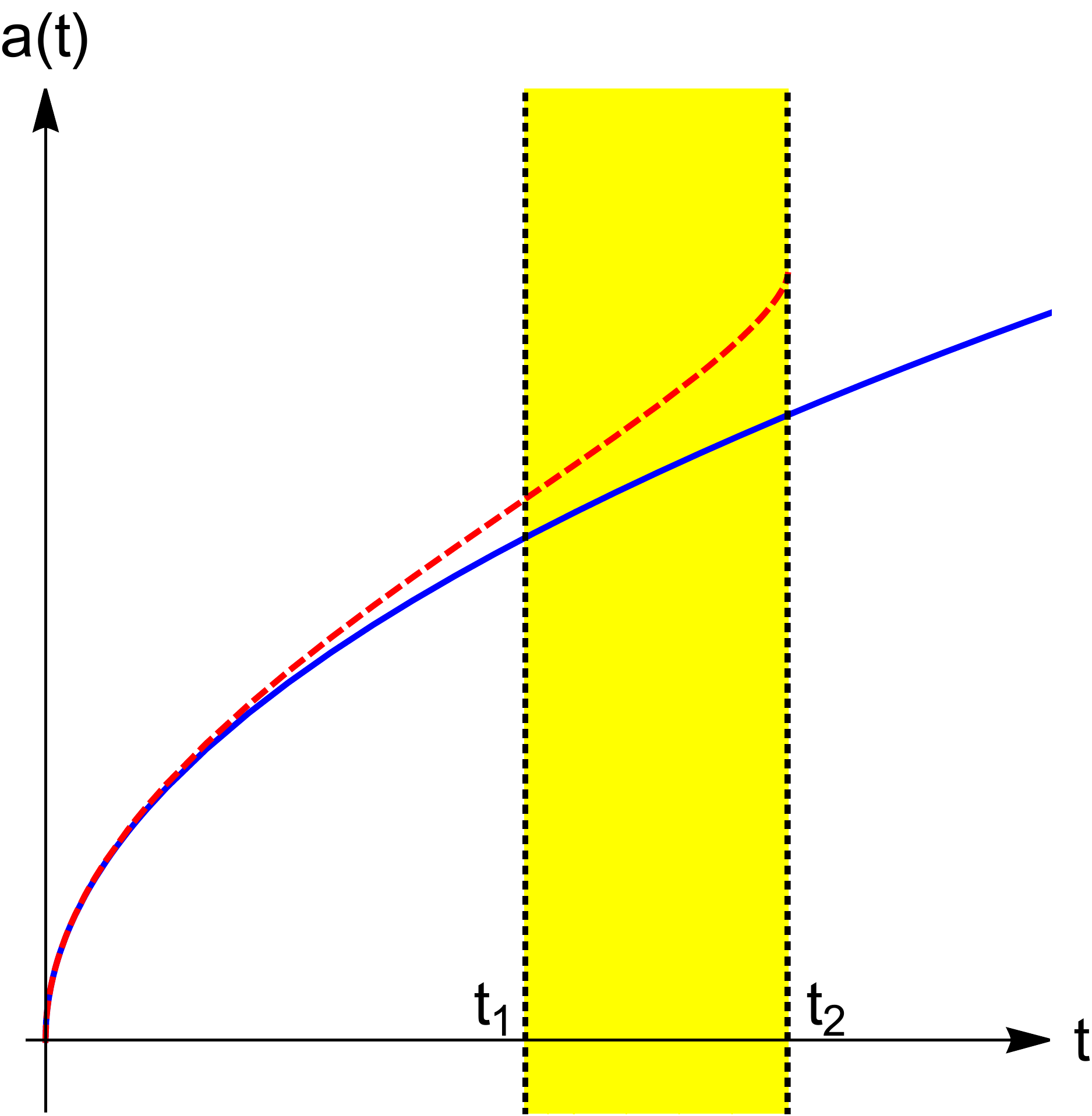}
\caption{Behavior of the scale factor as a function of time for $w=1/3$ and with the background of Eq.~\eqref{eq:sijnew} present. The blue (plain) line illustrates the standard behavior for $\alpha=0$ and the red (dashed) line shows the modified scale factor, which follows from solving Eq.~\eqref{eq:difeqa} numerically for $\alpha=9\times 10^{-3}$. The highlighted region indicates a time window of accelerated expansion. Furthermore, the black (dotted) vertical line on the left-hand side illustrates the instant of time $t_1$ starting from which Eq.~\eqref{eq:infws2} is satisfied. The vertical line on the right-hand side shows the instant of time $t_2$ where the modified solution becomes complex.}
\label{fig:numasolved}
\end{figure}

\begin{figure}
\centering
\includegraphics[scale=0.4]{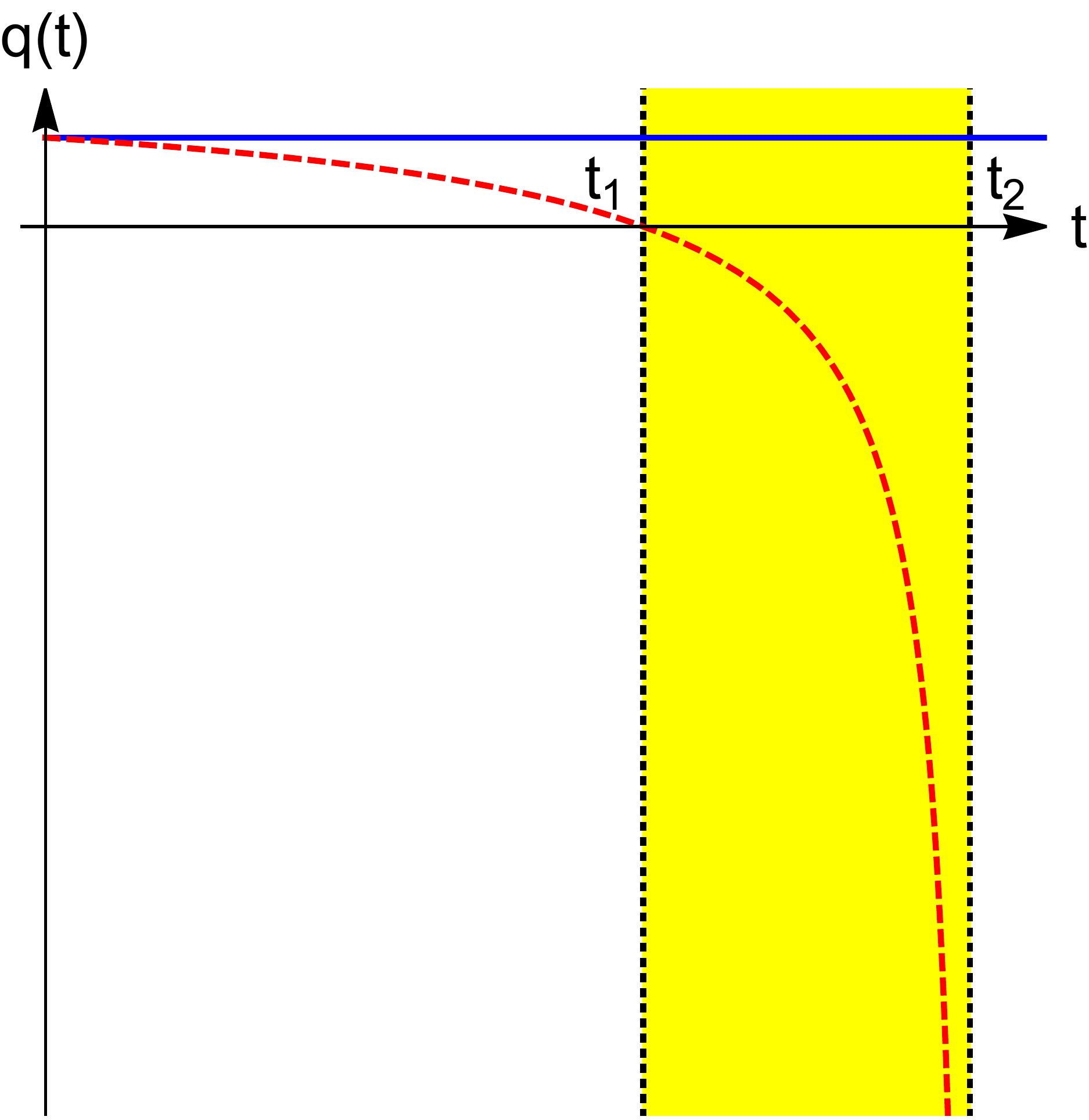}
\caption{Plot of the deceleration parameter $q(t)$ of Eq.~\eqref{eq:deceleration-parameter} obtained for the numerical solution of Eq.~\eqref{eq:difeqa} for $\alpha=9\times 10^{-3}$ where the blue, plain (red, dashed) curve shows the standard (modified) behavior. The highlighted region indicates the period $t\in [t_1,t_2]$ characterized by a $q(t)<0$. The existence of this region is clearly related to the range of the scale factor given in Eq.~\eqref{eq:range-scale-factor-radiation}.}
\label{fig:deceleration-parameter}
\end{figure}

An explicit computation of $\nabla_{\alpha}(T^{Rs})^{\alpha\beta}$ in Eq.~\eqref{eq:dtabs} shows that it is identically zero. Thus, this particular case is intriguing, as the no-go results \cite{Kostelecky:2003fs,Kostelecky:2020hbb} do not lead to further restrictions of the scale factor. Moreover, the choice of Eq.~\eqref{eq:sijnew} also satisfies $D_i D^i s = 0$. We then deduce from Eq.~\eqref{eq:inequalities} that
\begin{equation}
1 + 3 w < \frac{2 \alpha a^2}{1 - \alpha a^2} \Leftrightarrow \frac{1 + 3 w}{3 (1 + w) \alpha} < a^2\,.
\end{equation}
Hence, by taking Eq.~\eqref{eq:infws2} into account, accelerated expansion for standard matter with $w=0$ occurs when the scale factor squared lies within the range given by
\begin{subequations}
\begin{equation}
\frac{1}{3 \alpha} < a^2 < \frac{1}{\alpha}\,,
\end{equation}
where for standard radiation with $w=1/3$ it must hold that
\begin{equation}
\label{eq:range-scale-factor-radiation}
\frac{1}{2 \alpha} < a^2 < \frac{1}{\alpha}\,.
\end{equation}
\end{subequations}
Now let us turn to Eq.~\eqref{eq:friedmann-spatial}, which can be reformulated with the help of our previous findings:
\begin{subequations}
\begin{align}
\label{eq:frdmod}
H^2 &= \frac{1}{3 (1 - \alpha a^2)} \rho\,, \\[2ex]
\label{eq:frdmod2}
\dot{H} + H^2 &= - \frac{1}{6 (1 - \alpha a^2)} (\rho + 3 P - 6 \alpha a^2H^2)\,.
\end{align}
\end{subequations}
Note that Eq.~\eqref{eq:frdmod2} is a consequence of Eq.~\eqref{eq:frdmod} and energy-momentum conservation for matter, i.e., Eq.~\eqref{eq:rhowitha}. So after taking the no-go conditions into account, the second Friedmann results from the first such as in GR --- recall the statements made under Eq.~\eqref{eq:purely-spacelike-einstein-equations}. Hence, it is sufficient to study Eq.~\eqref{eq:frdmod}, which then takes the form
\begin{equation}
\label{eq:difeqa}
\dot{a} = \frac{\zeta}{\sqrt{1 - \alpha a^2}} a^{- \frac{1}{2} (1 + 3 w)}\,,\quad \zeta = \sqrt{\frac{\rho_0}{3}} (a_0)^{\frac{3}{2}(1 + w)}\,.
\end{equation}
We solve Eq.~\eqref{eq:difeqa} numerically by considering $\zeta=1$, $\alpha=9\times 10^{-3}$, and $w=1/3$ for a radiation-dominated Universe. The numerical solution for the scale factor is contrasted with the standard behavior of the latter in Fig.~\ref{fig:numasolved}.
Furthermore, Fig.~\ref{fig:deceleration-parameter} illustrates the deceleration parameter $q(t)$ of Eq.~\eqref{eq:deceleration-parameter} to show that the numerical solution satisfies $q<0$ in the region of interest.

Our observation is that the time frame where Eq.~\eqref{eq:infws2} is satisfied coincides with $\ddot{a}>0$ being valid, as expected. Hence, there is a short period where an accelerated expansion takes place and at the end of this stage the scale factor becomes imaginary. This result is expected in Eq.~\eqref{eq:difeqa}, because the square root is real only if $1-\alpha a^2>0$. However, when $t$ exceeds a certain instant of time, $a(t)$ becomes complex and loses its interpretation as a scale factor. This behavior is traced back to the requirement of satisfying Eq.~\eqref{eq:bound-s} without any sources of exotic matter or radiation present to drive the accelerated expansion.

\subsubsection{Angle-independent $s^{ij}$}

A second possible choice for a purely spacelike background tensor is
\begin{equation}
\label{eq:sijnew2}
s^{\mu \nu} = -2\frac{L^2}{r^2}\left(\begin{array}{cccc}
  0 & 0 & 0 & 0 \\
  0 & 1 & 0 & 0 \\
  0 & 0 & 0 & 0 \\
  0 & 0 & 0 & 0 \\
\end{array}\right)\,,
\end{equation}
with a length scale $L$ that must be introduced for dimensional consistency. Note that $s^{\mu\nu}$ is a dimensionless object --- at least in Cartesian coordinates. As $s^{\mu\nu}$ effectively incorporates Planck scale effects, the length scale $L$ is expected to lie in the vicinity of the Planck length. An explicit evaluation of Eq.~\eqref{eq:dtabs} gives a single component that is not automatically equal to zero:
\begin{equation}
\label{eq:no-go-s}
\nabla_{\alpha} T_s^{\alpha 1} = \frac{2L^2}{a^2r^3} (2 \dot{a}^2 + a \ddot{a})\,.
\end{equation}
Then, the no-go conditions of Eq.~\eqref{eq:no-go-constraint-einstein-equation-s} imply a restriction of the scale factor:
\begin{equation}
\label{eq:constraintsija}
2 \dot{a}^2 + a \ddot{a} = 0\,.
\end{equation}
The latter equation is associated with a constant value for the deceleration parameter of Eq.~\eqref{eq:deceleration-parameter}. This value amount to $q=2$, which is positive, whereupon the choice of Eq.~\eqref{eq:sijnew2} forbids a stage of accelerated expansion of the Universe.

\subsubsection{Peculiar choice of $s^{ij}$}\label{sec:raresij}

Finally, let us consider another choice of a purely spacelike background field $s^{\mu\nu}$ given by
\begin{equation}
\label{eq:invisiblesij}
s^{\mu \nu} =\frac{\tilde{L}^3}{r^4} \left(\begin{array}{cccc}
  0 & 0 & 0 & 0 \\
  0 & 0 & 0 & 1 \\
  0 & 0 & 0 & 0 \\
  0 & 1 & 0 & 0 \\
\end{array}\right)\,,
\end{equation}
with another length scale $\tilde{L}$ introduced for dimensional consistency. The latter choice leads to an interesting conclusion. Since $s=0$, it satisfies the relation of Eq.~\eqref{eq:constraintsij}. With this choice the Friedmann equations are not modified, at all. Furthermore, Eq.~\eqref{eq:no-go-constraint-einstein-equation-s} does not provide another constraint for the scale factor.
Therefore, a new source violating the SEC is definitely required to drive accelerated expansion. This finding teaches us that certain background fields such as Eq.~\eqref{eq:invisiblesij} may be present without having any impact on cosmological time evolution. They are deemed to be unobservable from a cosmological point of view.

\section{General modified Friedmann equations}
\label{sec:relaxu}

Insights gained in the recent paper \cite{Reyes:2022mvm} suggest that Eq.~\eqref{eq:consistency-conditions} is not necessary to establish a consistent dynamics of the theory based on Eq.~\eqref{eq:actionus}. Thus, it is worthwhile to understand the cosmological impact of Eq.~\eqref{eq:actionus} without these conditions taken into account. By doing so, the background fields proper are no longer static and are directly affected by the time evolution of the Universe. So this scenario is more generic than that studied previously. The first modified Friedmann equation follows from computing the functional derivative of Eq.~\eqref{eq:ADM-action-reformulated} for $N$ without incorporating Eq.~\eqref{eq:consistency-conditions}. The second is obtained from the purely spacelike part of the modified Einstein equations, as described in Sec.~\ref{sec:cosmo}. They are cast into the form
\begin{subequations}
\begin{widetext}
\begin{align}
\label{eq:friedmann-1-time-dependent-case}
H^2&=\frac{1}{3(1-u-s^{\mathbf{nn}}+s/3)}\left[\rho+\Lambda-3\left(1-u+\frac{s}{3}\right)\frac{k}{a^2}-D_iD^iu-\frac{1}{2}D_iD^is^{\mathbf{nn}}+\frac{1}{2}D_iD_js^{ij}\right. \notag \displaybreak[0]\\
&\phantom{{}={}}\hspace{3.4cm}\left.{}+3H\left(\dot{u}+\frac{\dot{s}^{\mathbf{nn}}}{2}-\frac{1}{6}\dot{s}^{ij}h_{ij}\right)\right]\,, \displaybreak[0]\\[2ex]
\label{eq:friedmann-2-time-dependent-case}
\dot{H}+H^2&=-\frac{1}{6(1-u-s^{\mathbf{nn}}+s/3)}\left(\rho+3P-2\Lambda+2s\left[\frac{k}{a^2}+H^2\right]+D_iD^iu-\frac{1}{2}D_iD^is^{\mathbf{nn}}-\frac{1}{2}D_iD^is\right. \notag \displaybreak[0]\\
&\phantom{{}={}}\hspace{3.8cm}\left.{}-3\left[H\left(\dot{u}+\frac{3}{2}\dot{s}^{\mathbf{nn}}-\frac{1}{2}\dot{s}^{ij}h_{ij}\right)+\ddot{u}+\frac{\ddot{s}^{\mathbf{nn}}}{2}-\frac{1}{6}\ddot{s}^{ij}h_{ij}\right]\right)\,.
\end{align}
\end{widetext}
\end{subequations}
The modified Friedmann equations for static background fields stated in Eq.~\eqref{eq:constraint1} and Eq.~\eqref{eq:secfriedmannb}, respectively, are recovered for $\dot{u}=\dot{s}^{\mathbf{nn}}=\dot{s}^{ij}=0$, as expected. In the scenario of time-dependent backgrounds, there are additional contributions that involve products of the Hubble parameter with time derivatives of background fields. Note that Eq.~\eqref{eq:friedmann-2-time-dependent-case} even contains second-order time derivatives of the backgrounds.

Furthermore, there is an additional constraint, which is inferred from the functional derivative of Eq.~\eqref{eq:ADM-action-reformulated} for~$N^i$:
\begin{align}
\label{eq:momentum-constraint-nonstatic-case}
0&=D_i\bigg[H(4+2u-s^{\mathbf{nn}}-s)h^{ik}+2Hs^{ik} \notag \\
&\phantom{{}={}}\hspace{0.5cm}-2h^{ik}\bigg(\dot{u}+\frac{\dot{s}^{\mathbf{nn}}}{2}\bigg)+\dot{s}^{ik}\bigg]\,.
\end{align}
The latter is automatically satisfied for vanishing controlling coefficients as was the case for Eq.~\eqref{eq:constraint2} derived earlier for static backgrounds. In what follows, we will again discard the cosmological constant $\Lambda$ as well as the scalar curvature $k$.

\subsection{Scalar background $u$}

We start by considering the scalar background field $u$ and set $s^{\mathbf{nn}}=s^{ij}=0$. Since $\mathcal{L}_mu$ is now taken to be different from zero, the modified Friedmann equations deduced from Eqs.~(\ref{eq:friedmann-1-time-dependent-case}), (\ref{eq:friedmann-2-time-dependent-case}) read
\begin{subequations}
\begin{align}
H^2&=\frac{1}{3(1-u)}(\rho-D_iD^iu+3H\dot{u})\,, \\[2ex]
\dot{H}+H^2&=-\frac{1}{6(1-u)}\left[\rho+3P+D_iD^iu-3(H\dot{u}+\ddot{u})\right]\,,
\end{align}
\end{subequations}
where both involve spatial as well as time derivatives of $u$. We must also bear in mind the additional constraint from Eq.~\eqref{eq:momentum-constraint-nonstatic-case}:
\begin{equation}
0 = D_i [H (4 + 2 u) h^{i k} - 2 h^{i k} \dot{u}]\,.
\end{equation}
The latter implies
\begin{equation}
\label{eq:momentum-constraint-u-reformulated}
0 = D_i (H u - \dot{u})\,.
\end{equation}
Moreover, the background field $u$ must obey the condition of Eq.~\eqref{eq:no-go-constraints-u}, i.e.,
\begin{equation}
\label{eq:no-go-constraint-u}
\nabla_\alpha u =\partial_{\alpha}u = 0\,.
\end{equation}
This requirement is only satisfied for constant $u$. As we already found in Sec.~\ref{sec:utime}, a constant $u$ can be eliminated by rescaling the matter density and pressure. Hence, novel, interesting cosmological scenarios based on a nonzero background field $u$ are not gained from discarding $\mathcal{L}_mu=0$.

\subsection{Purely timelike background $s^{\mathbf{nn}}$}

The next step is to study the second scalar background field $s^{\mathbf{nn}}$ with $u=s^{ij}=0$. In this case, Eqs.~(\ref{eq:friedmann-1-time-dependent-case}), (\ref{eq:friedmann-2-time-dependent-case}) imply
\begin{subequations}
\label{eq:modified-friedmann-equations-snn-1}
\begin{align}
H^2&=\frac{1}{3(1-s^{\mathbf{nn}})}\left(\rho-\frac{1}{2}D_iD^is^{\mathbf{nn}}+\frac{3}{2}H\dot{s}^{\mathbf{nn}}\right)\,, \\[2ex]
\dot{H}+H^2&=-\frac{1}{6(1-s^{\mathbf{nn}})}\left[\rho+3P-\frac{1}{2}D_iD^is^{\mathbf{nn}}\right. \notag \\
&\phantom{{}={}}\hspace{2cm}\left.{}-\frac{3}{2}(3H\dot{s}^{\mathbf{nn}}+\ddot{s}^{\mathbf{nn}})\right]\,.
\end{align}
\end{subequations}
The constraint of Eq.~\eqref{eq:momentum-constraint-nonstatic-case} provides
\begin{equation}
\label{eq:momentum-constraint-snn}
0 = D_i [H (4 - s^{\mathbf{nn}}) h^{i k} - h^{i k}\dot{s}^{\mathbf{nn}}]\,.
\end{equation}
Besides, the no-go restriction of Eq.~\eqref{eq:dtabs} applied to this particular case requires us to compute
\begin{subequations}
\begin{align}
\label{eq:no-go-condition-snn-component-00}
\nabla_{\alpha} (T^{Rs})^{\alpha 0} &= \frac{1}{2 a^2}\left[6 s^{\mathbf{nn}} \left( 2 \dot{a} \ddot{a} + a
\frac{\mathrm{d}^3 a}{\mathrm{d} t^3} \right) + 9 a \ddot{a} \dot{s}^{\mathbf{nn}}\right]\,, \displaybreak[0]\\[2ex]
\nabla_{\alpha} (T^{Rs})^{\alpha 1} &= - \frac{3 \ddot{a}}{2 a^3} \partial_r
s^{\mathbf{nn}}\,, \displaybreak[0]\\[2ex]
\nabla_{\alpha} (T^{Rs})^{\alpha 2} &= - \frac{3 \ddot{a}}{2 a^3 r^2}
\partial_{\theta} s^{\mathbf{nn}}\,, \displaybreak[0]\\[2ex]
\nabla_{\alpha} (T^{Rs})^{\alpha 3} &= - \frac{3 \ddot{a}}{2 a^3 r^2 \sin^2 \theta}
\partial_{\phi} s^{\mathbf{nn}}\,.
\end{align}
\end{subequations}
The condition $\nabla_{\alpha}(T^{Rs})^{\alpha\beta}=0$ for $\beta=1,2,3$ implies that $s^{\mathbf{nn}}$ can only depend on time. For a constant $s^{\mathbf{nn}}$, we recover the outcome of Sec.~\ref{sec:snntime}, as expected. Equation~\eqref{eq:no-go-condition-snn-component-00} further leads to
\begin{equation}
\frac{1}{2 a^2}\left[6 s^{\mathbf{nn}} \left( 2 \dot{a} \ddot{a} + a \frac{\mathrm{d}^3 a}{\mathrm{d} t^3} \right) + 9a \ddot{a} \dot{s}^{\mathbf{nn}}\right] = 0\,,
\end{equation}
which can be solved for the time derivative of the background field:
\begin{equation}
\label{eq:time-derivative-snn-no-go}
\dot{s}^{\mathbf{nn}} = - \frac{2}{3} \left( \frac{1}{\ddot{a}}  \frac{\mathrm{d}^3 a}{\mathrm{d} t^3} +
2 H \right) s^{\mathbf{nn}}\,.
\end{equation}
Aside from that, the extra constraint of Eq.~\eqref{eq:momentum-constraint-snn} provides $0 = \partial_i (H s^{\mathbf{nn}} + \dot{s}^{\mathbf{nn}})$. As a consequence, $H s^{\mathbf{nn}} + \dot{s}^{\mathbf{nn}} = f$ with a time-dependent function $f=f(t)$. Then the first-order time derivative of the background field amounts to
\begin{equation}
\label{eq:momentum-constraint-snn2}
\dot{s}^{\mathbf{nn}} = f - H s^{\mathbf{nn}}\,.
\end{equation}
By comparing Eq.~\eqref{eq:time-derivative-snn-no-go} with Eq.~\eqref{eq:momentum-constraint-snn2} we observe that $f = 0$ must be set to eliminate $s^{\mathbf{nn}}$ from both sides of the equation. So,
\begin{equation}
H = \frac{2}{3} \left( \frac{1}{\ddot{a}}  \frac{\mathrm{d}^3a}{\mathrm{d}t^3} + 2H \right)\,,
\end{equation}
which leads to the Hubble parameter:
\begin{equation}
H = - \frac{2}{\ddot{a}}\frac{\mathrm{d}^3 a}{\mathrm{d} t^3}\,.
\end{equation}
The latter has a solution of the form $a(t)=(t/t_0)^{4/3}$ where $t_0$ is an initial time. With this result at our disposal, we have
\begin{equation}
\label{eq:condforsnn}
H s^{\mathbf{nn}} + \dot{s}^{\mathbf{nn}} = 0\,,
\end{equation}
which allows us to determine $s^{\mathbf{nn}}=s^{\mathbf{nn}}(t)=1/a(t)$. Note that the previous information has solely been obtained from the no-go conditions and the constraint of Eq.~\eqref{eq:momentum-constraint-snn}.

Having determined the solutions for the scale factor $a(t)$ and the background field $s^{\mathbf{nn}}$, we can devote ourselves to the modified Friedmann equations, which have not been employed, so far. Taking into account that $s^{\mathbf{nn}}$ does not depend on the spatial coordinates, the modified Friedmann equations~\eqref{eq:modified-friedmann-equations-snn-1} read
\begin{subequations}
\begin{align}
\label{eq:friedmann-1-snn-nonstatic}
H^2 &= \frac{1}{3 (1 - s^{\mathbf{nn}})} \left( \rho +
\frac{3}{2} H \dot{s}^{\mathbf{nn}} \right)\,, \\[2ex]
\dot{H} + H^2 &= - \frac{1}{6 (1 - s^{\mathbf{nn}})} \left( \rho
+ 3 P - \frac{9}{2} H \dot{s}^{\mathbf{nn}} - \frac{3}{2}\ddot{s}^{\mathbf{nn}} \right)\,.
\end{align}
\end{subequations}
At this point, a comparison of the latter to Eqs.~(64), (65) of Ref.~\cite{ONeal-Ault:2020ebv} is worthwhile. A reasonable starting point is established by subtracting $1/2$ of their Eq.~(64) from Eq.~(65), which provides the combination $\dot{H}+H^2$ on the left-hand side. Then, it is intriguing to observe that their modified Friedmann equations almost correspond to ours in the $s^{\mathbf{nn}}$ sector except of the $s_{00}\ddot{a}/a$ term in their Eq.~(64), which does not have a counterpart in our Eq.~\eqref{eq:friedmann-1-snn-nonstatic}. However, one must take into account that the authors of Ref.~\cite{ONeal-Ault:2020ebv} study a modified-gravity theory where $R^{\mu\nu}$ is contracted with the lower-index background field $s_{\mu\nu}$. Furthermore, they employ a different methodology to derive their modified Friedmann equations.

Now, by using Eq.~\eqref{eq:condforsnn} and its derivative in the Friedmann equations, a little algebra allows them to be recast as follows:
\begin{subequations}
\begin{align}
H^2 &= \frac{2}{3 (2 - s^{\mathbf{nn}})} \rho\,, \label{eq:snnfriedmannarr} \displaybreak[0]\\[2ex]
\dot{H} + H^2 &= - \frac{2}{3 (4 - 3 s^{\mathbf{nn}})} \left( \rho + 3 P + \frac{3}{2} H^2 s^{\mathbf{nn}} \right)\,.
\end{align}
\end{subequations}
In Eq.~\eqref{eq:snnfriedmannarr} we employ Eq.~\eqref{eq:rhowitha}, which expresses the matter density in terms of the scale factor. Also, $s^{\mathbf{nn}}=s^{\mathbf{nn}}(t)=1/a(t)$ with $a(t)=(t/t_0)^{4/3}$ such that
\begin{equation}
2\left(\frac{t}{t_0}\right)^2 - \left(\frac{t}{t_0}\right)^{\frac{2}{3}} - \frac{3}{8}\rho_0a^{3 (1 + w)}_0t_0^2 \left(\frac{t}{t_0}\right)^{- 4 w} = 0\,.
\end{equation}
The latter algebraic equation is only satisfied for particular instants of time. Our conclusion is that the modified Friedmann equations for $s^{\mathbf{nn}}$ are incompatible.

\subsection{Tensor-valued purely spacelike background $s^{ij}$}

Finally, we study the scenario of a nonzero tensor-valued background field $s^{ij}$ and set $u=s^{\mathbf{nn}}=0$. Here, the modified Friedmann equations obtained from Eqs.~(\ref{eq:friedmann-1-time-dependent-case}), (\ref{eq:friedmann-2-time-dependent-case}) read
\begin{subequations}
\begin{align}
H^2&=\frac{1}{3(1+s/3)}\left(\rho+\frac{1}{2}D_iD_js^{ij}-\frac{H}{2}\dot{s}^{ij}h_{ij}\right)\,, \\[2ex]
\dot{H}+H^2&=-\frac{1}{6(1+s/3)}\left[\rho+3P-\frac{1}{2}D_iD^is+2sH^2\right. \notag \\
&\phantom{{}={}}\hspace{2cm}\left.{}+\frac{1}{2}(3H\dot{s}^{ij}h_{ij}+\ddot{s}^{ij}h_{ij})\right]\,.
\end{align}
\end{subequations}
The constraint following from Eq.~\eqref{eq:momentum-constraint-nonstatic-case} amounts to
\begin{equation}
0=D_i\left[H(4-s)h^{ik}+2Hs^{ik}+\dot{s}^{ik}\right]\,.
\end{equation}
Solving these equations together with the no-go results of Eq.~\eqref{eq:no-go-constraint-einstein-equation-s} has turned out to be highly challenging. Therefore, we leave it open as an interesting task to be tackled in future papers to come.

\section{Conclusions and outlook}\label{sec:con}

In this work we analyzed modified cosmologies based on particular nondynamical scalar- and tensor-valued background fields of the gravitational SME~\cite{Kostelecky:2003fs,Bailey:2006fd,Kostelecky:2017zob,Mewes:2019dhj,Kostelecky:2020hbb}. These backgrounds give rise to diffeomorphism violation, which we expected to have far-reaching consequences on the time evolution of the Universe. Our primary interest was to answer the question whether or not field configurations exist that are able to drive an accelerated expansion of the Universe with only standard matter and radiation present. The completely obscure nature of Dark Energy, which in the $\Lambda$CDM model is taken as the driving force behind the current accelerated expansion of the Universe, largely served as an incentive to do so. In addition, the results were expected to be equally applicable to inflation.

We focused on two background fields that are known as $u$ and $s^{\mu\nu}$ in the SME literature. The analysis was based on the modified Einstein equations as well as the ADM-decomposed action of the gravitational SME, which had been developed for these backgrounds in earlier works~\cite{Reyes:2021cpx,Reyes:2022mvm}. A decomposition of $s^{\mu\nu}$ into a purely timelike sector governed by a single coefficient $s^{\mathbf{nn}}$ and a purely spacelike sector parameterized by six coefficients $s^{ij}$ turned out to be convenient. In the first part of the investigation, each background field was assumed to be independent of time. Using this additional restriction simplified the computations and was a first step to understand the implications of the backgrounds $u$, $s^{\mathbf{nn}}$, and $s^{ij}$ on cosmology. However, this restriction was entirely dropped in the second part of the paper, introducing further technical complications.

Moreover, we took into consideration additional no-go conditions that emerge in the presence of nondynamical background fields in modified-gravity theories~\cite{Kostelecky:2003fs,Kostelecky:2020hbb,Bluhm:2014oua,Bluhm:2016dzm}. We then derived the first and second modified Friedmann equations for each of the sectors governed by $u$, $s^{\mathbf{nn}}$, and $s^{ij}$. On the one hand, the scalar background field $u$ was found to act as a mere scaling factor that can be absorbed into a redefined density and pressure of standard matter. Therefore, it does not imply a nonstandard regime of accelerated expansion of the Universe. Furthermore, although we assert that $s^{\mathbf{nn}}$ is not a simple scaling factor, it does also not result in a stage of accelerated expansion.

On the other hand, the time evolution of the Universe based on a nonzero $s^{ij}$ was found to exhibit more interesting behaviors. We studied different explicit choices of this background. One choice implies an accelerated expansion without exotic matter present, but only within a very restricted time frame. At a certain instant of time, the scale factor becomes complex and loses its physical interpretation, which is indicative of a breakdown of the effective model under study. Finding other background field configurations that lead to the desired behavior turned out to be intricate. Some configurations were encountered to reproduce the standard evolution equations, i.e., they are not expected to be observable in cosmology, at all.

The outcomes of this paper demonstrate the challenge of finding suitable choices for nontrivial background fields $u$ and $s^{\mu\nu}$ that satisfy all the necessary requirements of Eqs.~\eqref{eq:consistency-conditions}, \eqref{eq:no-go-constraints-u}, \eqref{eq:no-go-constraint-einstein-equation-s} and provide interesting cosmological behaviors. Unfortunately, as was demonstrated in Sec.~\ref{sec:relaxu}, the situation does not change much by dropping Eq.~\eqref{eq:consistency-conditions}, i.e., by allowing for time-dependent background fields. There is still too little freedom in choosing background fields $u$ and $s^{\mu\nu}$ such that an accelerated expansion can occur over long periods without exotic forms of matter and radiation. To the best of our knowledge, together with Refs.~\cite{Bonder:2017dpb,ONeal-Ault:2020ebv,Nilsson:2022mzq}, our paper is one of the first to apply the gravitational SME to cosmology. The methods developed and findings made shall serve as a precursor for further research to be carried out in this interesting subfield.

One possibility of disregarding the very restrictive no-go results of Eqs.~\eqref{eq:no-go-constraints-u}, \eqref{eq:no-go-constraint-einstein-equation-s} would be to elaborate modified-gravity theories based on the gravitational SME with diffeomorphism invariance violated spontaneously such as in bumblebee-type models \cite{Bluhm:2004ep,Bluhm:2007bd,Bluhm:2008yt,Seifert:2009vr,Seifert:2009gi,Hernaski:2014jsa,Bonder:2015jra,Maluf:2015hda,Casana:2017jkc,Colladay:2019lig,Assuncao:2019azw,Maluf:2020kgf,Poulis:2021nqh,Delhom:2022xfo}. By doing so, conflicts with the Bianchi identities of Riemannian geometry are neatly avoided. Other complications may then arise by having to choose suitable potentials for the background fields as well as having to take fluctuations of the backgrounds into account. The latter transform in a nontrivial way under diffeomorphisms, i.e., they are indispensable to restore diffeomorphism symmetry of the theory.

Another intriguing question to answer could be whether the method of functional derivatives of the ADM-decomposed action provides the same modified Friedmann equations as the Hamiltonian or covariant approaches. Based on the findings of Ref.~\cite{Reyes:2021cpx}, this is definitely expected to be the case for background fields that obey Eq.~\eqref{eq:consistency-conditions}. According to Ref.~\cite{Reyes:2022mvm}, the Hamiltonian approach is consistent with the covariant formalism even when dropping Eq.~\eqref{eq:consistency-conditions}. However, what can be said about the technique that relies on the functional derivatives of the action is open for investigation. These lines of research may provide a worthwhile future continuation of the current project.

\begin{acknowledgements}

C.M.R acknowledges partial support by the research project Fondecyt Regular 1191553. M.S. is indebted to FAPEMA Universal 00830/19, CNPq Produtividade 310076/2021-8, and CAPES/Finance Code 001.

\end{acknowledgements}

\appendix

\begin{appendix}

\section{Alternative evolution equations}
\label{sec:alternative-equations}

Here we intend to derive another set of equations from the first modified Friedmann equation by following the procedure often employed in GR. First of all, we will focus on time-independent background fields. By differentiating Eq.~\eqref{eq:constraint1} for the time variable, using Eq.~\eqref{eq:continuity}, and inserting the first Friedmann equation again, we arrive at an alternative modification of the second Friedmann equation of GR. There are two possibilities of putting the latter on paper. The first is to have it depend on the scalar curvature $k$ explicitly. In the second form, $k$ is eliminated via the first Friedmann equation of Eq.~\eqref{eq:constraint1}:
\begin{align}
\frac{k}{a^2}&=\frac{1}{\Upsilon}\left[\frac{\rho}{3}+\frac{1}{6}(D_i D_j s^{i j} - D_i D^i s^{\mathbf{nn}} - 2 D_i D^i u)\right. \notag \\
&\phantom{{}={}}\hspace{0.5cm}\left.{}+\frac{\Lambda}{3}-\Xi H^2\right]\,.
\end{align}
The two equivalent forms are then given by
\begin{widetext}
\begin{subequations}
\begin{align}
\label{eq:friedmann-equation-2a}
\dot{H} + H^2 &\approx - \frac{1}{6\Xi} \bigg\{ \left( 1 + \frac{2 s}{3\Xi} \right) \rho + 3 P-\frac{2}{\Xi}(1-u-s^{\mathbf{nn}})\Lambda \notag \\
&\phantom{{}={}}\hspace{1cm}-\left[ 1 - \frac{s}{3\Upsilon}\left(1+\frac{s^{\mathbf{nn}}}{\Xi}\right) \right](D_iD_js^{ij}-D_iD^is^{\mathbf{nn}}-2D_iD^i u)-\frac{2ss^{\mathbf{nn}}}{\Xi}\frac{k}{a^2}\bigg\}\,,
\end{align}
and
\begin{align}
\label{eq:friedmann-equation-2b}
\dot{H}+H^2&\approx -\frac{1}{6\Xi}\bigg[\left(1+\frac{2s}{3\Upsilon}\right)\varrho+3P-2\Lambda-\left(1-\frac{s}{3\Upsilon}\right)(D_iD_js^{ij}-D_iD^is^{\mathbf{nn}}-2D_iD^i u)\bigg] \notag \\
&\phantom{{}={}}-\frac{s}{3\Xi\Upsilon}\left(s^{\mathbf{nn}}H^2+\frac{\Lambda}{3}\right)\,,
\end{align}
\end{subequations}
\end{widetext}
respectively. Here we employed the symbol $\approx$, which stands for ``weakly equal to zero'' \cite{Hanson:1976} in this context, as the treatment and interpretation of the previous relationships require special care. Since Eqs.~(\ref{eq:friedmann-equation-2a}), (\ref{eq:friedmann-equation-2b}) are deduced from the first modified Friedmann equation, which is a constraint, the latter equations are only valid whenever the original constraint is satisfied.

The number of terms in Eq.~\eqref{eq:friedmann-equation-2a} is larger compared to Eq.~\eqref{eq:friedmann-equation-2b}. However, the right-hand side of Eq.~\eqref{eq:friedmann-equation-2b} contains the Hubble parameter being not the case in Eq.~\eqref{eq:friedmann-equation-2a}. By comparing the latter to Eq.~\eqref{eq:secfriedmannb}, vast differences between these equations are evident. Note that Eq.~\eqref{eq:secfriedmannb} is linear in the background fields, whereas Eqs.~\eqref{eq:friedmann-equation-2b}, \eqref{eq:friedmann-equation-2b} are not.

Allowing for time-dependent background fields leads to
\begin{widetext}
\begin{align}
\label{eq:friedmann-equation-2-time-dependent-backgrounds}
\dot{H} + H^2&\approx -\frac{1}{6\Xi}\bigg\{\bigg( 1 + \frac{\dot{\Upsilon}}{H\Upsilon} \bigg)(\rho + \Lambda) + 3(P-\Lambda)- \bigg( 1 - \frac{\dot{\Upsilon}}{2H\Upsilon} \bigg) (D_i D_js^{ij}-D_iD^is^{\mathbf{nn}}-2D_iD^iu) \notag \displaybreak[0]\\
&\phantom{{}={}} - \frac{1}{2 H}(D_iD_j\dot{s}^{ij}-D_iD^i\dot{s}^{\mathbf{nn}}-2D_iD^i\dot{u})+2ss^{\mathbf{nn}}\frac{H^2}{\Upsilon}+\frac{\dot{H}}{2H}\left[\dot{s}^{ij}h_{ij}-3(\dot{u}+\dot{s}^{\mathbf{nn}})\right] \notag \displaybreak[0]\\
&\phantom{{}={}} + 3\frac{H}{\Upsilon}\left[(2-2u+s^{\mathbf{nn}})\dot{u}+\left(2-2u+\frac{s}{3}\right)\dot{s}^{\mathbf{nn}}\right] \notag \displaybreak[0]\\
&\phantom{{}={}} + \frac{1}{\Upsilon}\dot{s}^{ij}h_{ij}\left[-\frac{1}{6}\dot{s}^{kl}h_{kl}+H\left(2-2u+s^{\mathbf{nn}}+\frac{s}{3}\right)+\frac{3}{2}\left(\dot{u}+\frac{\dot{s}^{\mathbf{nn}}}{3}\right)\right] \notag \displaybreak[0]\\
&\phantom{{}={}}+\frac{1}{2}\ddot{s}^{ij}h_{ij}-\frac{3}{2}\left[\frac{\dot{u}}{\Upsilon}\left(2\dot{u}+\dot{s}^{\mathbf{nn}}\right)+2\ddot{u}+\ddot{s}^{\mathbf{nn}}\right]\bigg\}\,.
\end{align}
\end{widetext}
Here we also quickly observe that Eq.~\eqref{eq:friedmann-equation-2-time-dependent-backgrounds} strongly differs from Eq.~\eqref{eq:friedmann-2-time-dependent-case}, although each reproduces the second Friedmann equation of GR for vanishing controlling coefficients $u,s^{\mathbf{nn}},s^{ij}$. Furthermore, Eq.~\eqref{eq:friedmann-equation-2b} is reproduced from Eq.~\eqref{eq:friedmann-equation-2-time-dependent-backgrounds} when the background field coefficients are assumed to be static. A more sophisticated understanding of Eqs.~\eqref{eq:friedmann-equation-2a}, \eqref{eq:friedmann-equation-2b}, and \eqref{eq:friedmann-equation-2-time-dependent-backgrounds} requires an elaborate study of the constraint structure of Eq.~\eqref{eq:actionus} as well as the time evolution of constraints, which is beyond the scope of this work.

\end{appendix}


\end{document}